\documentclass[aip,jmp,a4paper,byrevtex,12pt,showpacs,showkeys,amsmath,amssymb,amsfonts,preprintnumbers,final,numerical]{revtex4}

\usepackage{graphicx}
\usepackage[all]{xy}

\begin{document}

\title{Revisitation of the original hot QCD collinear singularity problem}
\author{K. Bouakaz}
\affiliation{ Ecole Normale Sup\'erieure de Kouba, Alg\'erie}
\author{T. Grandou}
\affiliation{Institut NonLin\'eaire de Nice, 1361 route des Lucioles, Sophia Antipolis, 06560 Valbonne, France}

\begin{abstract}
The long standing issue known as the hot QCD collinear singularity problem has
been proven to rely on an incorrect sequence of two mathematical operations.
Here, the original derivation of this problem is entirely revisited within the correct sequence, bringing to light new and unexpected conclusions.
\end{abstract}


\maketitle

\section{Introduction}

The intrinsic non perturbative nature of non zero temperature Quantum field
Theories has been recognized for long \cite{landsman:91} . Naive thermal
perturbation theory can nevertheless be devised, both in imaginary and real
time formalisms \cite{[2]}, but then, it promptly appears that, under
certain circumstances, the original perturbative series must be
re-organized. Such an example of re-organization is provided by the so
called \textit{Resummation Program} \cite{[3]}. This program, $RP$ for
short, is a resummation scheme of the leading order thermal fluctuations
which, in the literature, are known under the spell of \textit{Hard Thermal
Loops}. Whenever one is calculating a physical process related to thermal
Green's functions whose external/internal legs are \textit{soft}, it is
mandatory to trade the naive thermal perturbation theory for the $RP$. The
softness alluded to above, refers to momenta on the order of the soft scale $%
gT$, where $T$, the temperature, stands for the \textit{hard} scale and $g$
for any relevant (bare/renormalized) and small enough coupling constant.

The $RP$ which has been set up in order to remedy an obvious lack
of completeness of the naive thermal perturbation theory, has produced
interesting, gauge-invariant results. It is true, however, that it has also
met difficulties in the infrared regime of the theories \cite{[4], [5]}. One
of these difficulties is the sixteen years old hot QCD collinear singularity
problem: When one calculates the soft photon emission rate out of a
Quark-Gluon Plasma (QGP) at thermal equilibrium, the Resummation Program is
in order, but it delivers an answer which is plagued with a collinear
singularity \cite{[6]}.

The hot QCD collinear singularity problem has long been quoted an
important issue of the Quantum Field Theory non-zero temperature context,
and only two years after its discovery, had already become textbook material 
\cite{[6]}. It is recorded as a serious obstruction to the high temperature
effective Perturbation Theory, and has motivated several attempts of
solution \cite{[5], [7], [8]}.

Among the solutions that have been proposed, the latter, in
Ref.8, has been adopted widely: It relies on a gauge-invariant introduction
of so-called \textit{thermal asymptotic masses} in either
bosonic and fermionic sectors of the theories. The thermal asymptotic masses
are on the order of the soft scale, $gT$, and as any mass, they are expected to 
\textit{screen} the logarithmic collinear singularity under consideration.

However, despite the fact that the introduction of such masses
suffers from a lack of justification, the singularity screening it provides
reveals itself not efficient enough beyond the stage of a one loop
calculation, and indeed, the problem bounces back.

This fate is due to the mechanism of \textit{collinear enhancement%
}, able to render higher number of loop contributions as important, if no
more important than lower number of loop calculations, \cite{[9]}. Needless
to say that in such a dramatic situation, the Resummation Program comes out
deprived of any reliability and predictive power, and that, by the time of
F. Gelis's thesis striking results, \cite{[9]}, people involved in the
matter were almost driven to despair. 

This has lead some authors to explore the possibility that extra topologies
of graphs be considered that could compensate for the logarithmic collinear
singularity of the original diagrams \cite{[10]}; and one of us, to get back
to the original derivation of the hot QCD collinear singularity problem \cite%
{[11]}.

In this latter instance, \cite{[11]}, it was discovered that the
collinear singularity original derivation hinged upon an incorrect sequence
of two mathematical steps to be taken, namely, an angular integration
followed by a prescription of discontinuity, to proceed along the correct
sequence.

In particular, it could be proven that the diagrams involving 1- 
\textit{effective soft photon-quark-anti-quark vertex}, the other one bare,
came out a regular quantity, in contradistinction to the original derivation
where the incorrect sequence was followed: Of course, this could be taken
as a serious invitation to revisit the whole problem within the correct
sequence.

Unfortunately, the much more involved 2-effective vertex diagram
remained an issue because of the incredibly long and difficult entwined
angular integrations it entails. However, that issue was the more decisive
as the original collinear singularity was explicitly due to that very
diagram and to no other. In other words, so long as the 2-effective vertex
diagram was not thoroughly calculated within the correct sequence, the hot
QCD collinear singularity problem could not be considered a fixed one.

Fixing definitely that issue is the task which is achieved in the
present article. The article
is organized as follows. Section 2 is a short reminder of the collinear
singularity problem met in hot QCD. This section will also serve the purpose
of introducing the quantities of interest as well as our notations. In
Section 3, the kinematics and the general structure of the calculations
involving $1$- and $2$- effective vertex diagrams are set up.

In order to reach sound conclusions, meticulous calculations of $%
1 $- and $2$-effective vertex diagrams must be carried out. Such is the case
of the former in Section 4, and of the latter in Section 5. Eventually, our
conclusions are drawn in Section 6, whereas two appendices complete the article. 

Throughout the article, we
will be using the convention of upper case letters for quadrimomenta and
lower case ones for their components, writing, for example $P\!=\!(p_{0},{%
\vec{p}})$. Our conventions for labelling internal and external momenta can
be read off Figure~\ref{fig:fig1}.

\section{The collinear singularity problem of hot $QCD$}

This sixteen years old issue is the following. The soft real photon
emission rate out of a Quark-Gluon Plasma in thermal equilibrium involves
the calculation of the quantity 
\[
\displaylines{\Pi_R(Q)=i\int {{\rm d}^4P\over
(2\pi)^4}(1-2n_F(p_0))\ {\rm
disc_P}\ Tr\biggl\lbrace {}^\star S_R(P)\  {}^\star\Gamma_\mu(P_R,Q_R,-P'_A)\cr\hfill {}^\star S_R
(P')\ {}^\star\Gamma^\mu(P_R,Q_R,-P'_A)\biggr\rbrace\qquad(2.1)} 
\]%
The discontinuity is to be taken in the energy variable $p_{0}$, by forming
the difference of $R$ and $A$-indiced $P$-dependent quantities, and within
standard notations, fermionic \textit{HTL} self energies, effective
propagators and vertices are respectively given by

$$
{}^\star S_\alpha(P)={\frac{i}{{\rlap / \!P} -
\Sigma_\alpha(P)+i\epsilon_\alpha p_0}}\ ,\ \ \ \ \alpha=R,A\ ,\ \ \ \
\epsilon_R=-\epsilon_A=\epsilon \eqno(2.2) 
$$

$$
\Sigma_\alpha(P)=m^2\int {\frac{\mathrm{d}{\widehat K}}{4\pi}} {\frac{{%
\rlap
/ \!\widehat K} }{{\ \widehat K}\!\cdot\! P+i\epsilon_\alpha}}\ ,\ \ \ \
m^2= C_F{\frac{g^2T^2}{8}}\eqno(2.3) 
$$

$$
{}^\star\Gamma_\mu(P_\alpha,Q_\beta,P^{\prime}_\delta)=-ie\left(\gamma_\mu+%
\Gamma^{HTL}_\mu(P_\alpha,Q_\beta,P^{\prime}_\delta)\right)\eqno(2.4) 
$$

$$
\Gamma _{\mu }^{HTL}(P_{\alpha },Q_{\beta },P_{\delta }^{\prime })=m^{2}\int 
{\frac{\mathrm{d}{\widehat{K}}}{4\pi }}{\frac{{\widehat{k}}_{\mu }\ {{\rlap /%
\!\widehat{K}}}}{({\widehat{K}}\!\cdot \!P+i\epsilon _{\alpha })({\widehat{K}%
}\!\cdot \!P^{\prime }+i\epsilon _{\delta })}}\eqno(2.5) 
$$%
where ${\widehat{K}}$ is the lightlike four vector $(1,{\widehat{k}})$. As
(2.4) is plugged into (2.1), four terms come about, three of them
proportional to a collinear singularity. These singular terms are the two
terms with one bare vertex $\gamma _{\mu }$, the other $\Gamma _{\mu }^{HTL}$%
, plus the term including two $HTL$ vertices, $\Gamma _{\mu }^{HTL}$. Thanks
to an abelian Ward identity peculiar to the high temperature limit, a
partial cancellation of these collinear singularities occurs, but out of the
term including two $\Gamma _{\mu }^{HTL}$ vertices, a collinear singularity
remains, 
\[
\displaylines {-{2i{e^2m^2\over q^2}}\left( \int{{\rm d}{\widehat
K}\over 4\pi}{1\over {\widehat
Q}\!\cdot\! {\widehat
K}+i\epsilon}\right)
\int {{\rm d}^4P\over
(2\pi)^3}\ \delta(P\!\cdot\!{\widehat Q})\ (1-2n_F(p_0))
\cr\hfill \times\  
[Tr\left({}^\star S_A(P){{\rlap / \!\widehat Q}}\right)-Tr\left({}^\star S_R(P'){{\rlap / \!\widehat
Q}}\right)]\qquad(2.6) } 
\]%
where, the soft photon being real, $Q$ is the lightlike $4$-vector $Q\!=q{%
\widehat{Q}}=\!q(1,{\widehat{q}})$, with $q$ a real positive number. In the
literature, this result is ordinarily written in the form 
$$
{\frac{C^{st}}{\varepsilon }}\int {\frac{{d^{4}P}}{(2\pi )^{4}}}\ \delta ({%
\widehat{Q}}\!\cdot \!P)\ (1-2n_{F}(p_{0}))\sum_{s=\pm 1,V=P,P^{\prime }}\pi
(1-s{\frac{v_{0}}{v}})\beta _{s}(V)\eqno(2.7) 
$$%
where the overall $1/\varepsilon $ comes from a dimensionally regularized
evaluation of the factored out angular integration appearing in (2.6), and
where $\beta _{s}(V)$ is related to the effective fermionic propagator usual
parametrization \cite{[6]}, 
$$
{}^{\star }S(P)={\frac{i}{2}}\sum_{s=\pm 1}{{{\rlap /\!\widehat{P}_{s}}}}%
{}^{\star }\Delta ^{s}(p_{0},p)\eqno(2.8) 
$$%
where ${\widehat{P_{s}}}=(1,s{\widehat{p}})$, the label $s$ referring to the
two dressed fermion propagating modes. One has 
$$
{}^{\star }\Delta ^{s}(p_{0},p)=\left( p_{0}-sp-{\frac{m^{2}}{2p}}[(1-s{%
\frac{p_{0}}{p}})\ln {\frac{p_{0}+p}{p_{0}-p}}+2s]\right) ^{-1}\eqno(2.9) 
$$%
the two Retarded/Advanced solutions corresponding to 
$$
{}^{\star }\Delta _{\alpha }^{s}(p_{0},p)\equiv {}^{\star }\Delta
^{s}(p_{0}+i\epsilon _{\alpha },p)=\alpha _{s}(p_{0},p)-i\pi \epsilon
(\epsilon _{\alpha })\beta _{s}(p_{0},p),\ \ \epsilon _{R}=-\epsilon _{A}=1%
\eqno(2.10) 
$$%
where $\epsilon (x)$ is the distribution "sign of $x$", and $\alpha =R,A$.

\section{$1$- and $2$-effective vertex contributions and
kinematics}

\begin{figure}
 \caption{Self energy diagram involving two effective propagators and one
vertex HTL correction}
\label{fig:fig1}
 \begin{center}
   \includegraphics[width=0.5\linewidth]{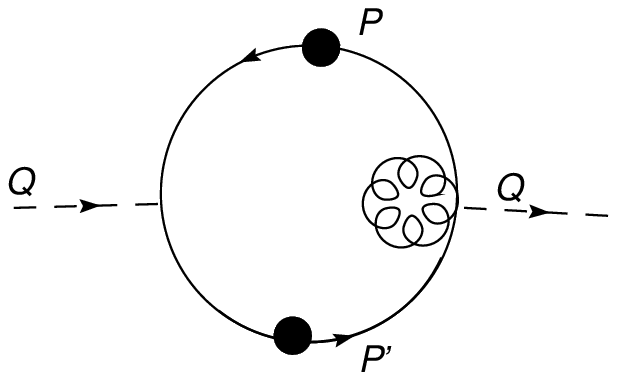}
 \end{center}
\end{figure}

The historical derivation just reminded above, however, is plagued
with erroneous manipulations that have been put forth in Ref.11. In the R/A
formalism, the two diagrams including one bare vertex $\gamma _{\mu }$, the
other $\Gamma _{\mu }^{HTL}$, (2.5), lead to the expression 
$$%
 \begin{aligned}
\Pi^{(\star,\star;1)}_R(Q) = -&ie^2m^2\int {{\rm d}^4P\over
(2\pi)^4}(1-2n_F(p_0)) \\
& {\rm
{disc}}_{p_0}\int{{\rm d}{\widehat
K}\over 4\pi}   {\ Tr\left({}^\star S_R(P){{\rlap / \!\widehat K}} {}^\star S_R(P'){{\rlap / \!\widehat
K}}\right)\over
({\widehat K}\!\cdot\! P+i\epsilon)({\widehat
K}\!\cdot\! P'+i\epsilon)}
 \end{aligned}
 \eqno(3.1)
$$%
where the superscript $(\star ,\star ;1)$ in the left hand side refers to a
self energy diagram involving two effective propagators and one vertex $HTL$
correction, as depicted in Figure~\ref{fig:fig1}. 

It is
this diagram which is now being analyzed within the correct sequence of
Eq.(3.1), where the angular average is to be performed before the
discontinuity in $p_{0}$ is taken. \smallskip \noindent One gets 
\[
\displaylines{ \Pi^{(\star,\star;1)}_R(Q)= 2ie^2m^2\int {{\rm d}^4P\over
(2\pi)^4}(1-2n_F(p_0))\cr\hfill\sum_{s,s'=\pm 1} {}^\star\Delta_R^{s'}(P')\biggl\lbrace -2i\pi\beta_s(P) \int{{\rm
d}{\widehat K}\over 4\pi} {{\widehat K}\!\cdot\!{\widehat P}_{s}\over {\widehat K}\!\cdot\!{\widehat
P}+i\epsilon}{{\widehat K}\!\cdot\!{\widehat P}'_{s'}\over {\widehat K}\!\cdot\!{\widehat
P'}+i\epsilon}\cr\hfill
+{}^\star\Delta_R^{s}(P){\rm{disc_{p_0}}}\int{{\rm d}{\widehat K}\over 4\pi} {{\widehat K}\!\cdot\!{\widehat
P}_{s}\over {\widehat K}\!\cdot\!{\widehat P}+i\epsilon}{{\widehat K}\!\cdot\!{\widehat P}'_{s'}\over {\widehat
K}\!\cdot\!{\widehat P'}+i\epsilon}\biggr\rbrace\qquad(3.2)}
\]%
where we have used $\mathrm{disc}_{p_{0}}{}^{\star }\Delta
_{R}^{s}(P)=-2i\pi \beta _{s}(P)$, whereas a factor of $2$ accounts for the
two $1$-effective vertex diagrams, which contribute equally. \medskip
Defining $W^{(1)}(P,P^{\prime })$ the function 
$$
W^{(1)}(P,P^{\prime })=\int {\frac{\mathrm{d}{\widehat{K}}}{4\pi }}{\frac{{%
\widehat{K}}\!\cdot \!{\widehat{P}}_{s}}{{\widehat{K}}\!\cdot \!{\widehat{P}}%
+i\epsilon }}{\frac{{\widehat{K}}\!\cdot \!{\widehat{P}}_{s^{\prime
}}^{\prime }}{{\widehat{K}}\!\cdot \!{\widehat{P}^{\prime }}+i\epsilon }}%
\eqno(3.3)
$$%
one obtains for the imaginary part, the expression 
\[
\displaylines{ {\rm{Im}}\  \Pi^{(\star,\star;1)}_R(Q)=2\pi e^2m^2\int {{\rm d}^4P\over
(2\pi)^4}(1-2n_F(p_0))\sum_{s,s'=\pm 1}\ \biggl\lbrace -2\pi\beta_s(P)\beta_{s'}(P')W^{(1)}(P,P')\cr\hfill 
+\left(\alpha_s(P)\beta_{s'}(P')+\alpha_{s'}(P')\beta_s(P)\right)\left(-i\ {\rm
{disc}}_{p_0}W^{(1)}(P,P')\right)\biggr\rbrace\qquad(3.4)}
\]

\begin{figure}
 \caption{Self energy diagram involving two effective propagators and two
 vertex HTL corrections}
\label{fig:fig2}
 \begin{center}
   \includegraphics[width=0.5\linewidth]{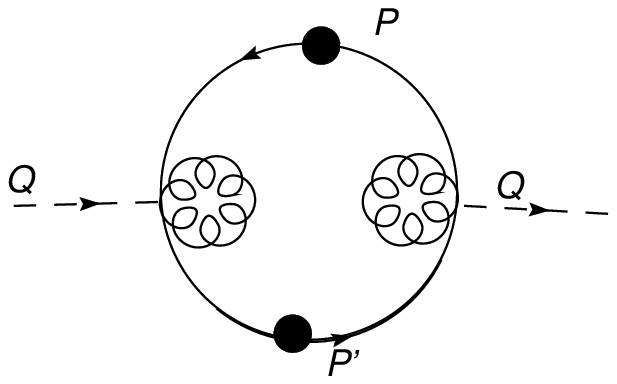}
 \end{center}
\end{figure}

\noindent For the diagram of Figure~\ref{fig:fig2}, involving two effective vertex,%

\[
\displaylines{\Pi^{(\star,\star;2)}_R(Q)=-ie^2m^4\int {{\rm d}^4P\over
(2\pi)^4}(1-2n_F(p_0))\cr\hfill
{\rm
{disc}}_{p_0}\int{{\rm d}{\widehat
K}\over 4\pi}\int{{\rm d}{\widehat
K'}\over 4\pi}{\widehat K}\!\cdot\! {\widehat K'} {\ Tr\left({}^\star S_R(P){{\rlap / \!\widehat K}} {}^\star S_R(P'){{\rlap / \!\widehat
K'}}\right)\over
({\widehat K}\!\cdot\! P+i\epsilon)({\widehat
K}\!\cdot\! P'+i\epsilon)({\widehat K'}\!\cdot\! P+i\epsilon)({\widehat
K'}\!\cdot\! P'+i\epsilon)}\qquad(3.5)}
\]%
so that, defining $W^{(2)}(P,P^{\prime })$ the function 
$$
W^{(2)}(P,P^{\prime })=\int {\frac{\mathrm{d}{\widehat{K}}}{4\pi }}\int {%
\frac{\mathrm{d}{\widehat{K}^{\prime }}}{4\pi }}\ {\widehat{K}}\!\cdot \!{%
\widehat{K}^{\prime }}{\frac{{\widehat{K}}\!\cdot \!{\widehat{P}}_{s}\ {%
\widehat{K}^{\prime }}\!\cdot \!{\widehat{P}^{\prime }}_{s^{\prime }}+{%
\widehat{K}}\!\cdot \!{\widehat{P}^{\prime }}_{s^{\prime }}\ {\widehat{K}%
^{\prime }}\!\cdot \!{\widehat{P}}_{s}-{\widehat{K}}\!\cdot \!{\widehat{K}%
^{\prime }}{\widehat{P}}_{s}\!\cdot \!{\widehat{P}^{\prime }}_{s^{\prime }}}{%
({\widehat{K}}\!\cdot \!P+i\epsilon )({\widehat{K}}\!\cdot \!P^{\prime
}+i\epsilon )({\widehat{K}^{\prime }}\!\cdot \!P+i\epsilon )({\widehat{K}%
^{\prime }}\!\cdot \!P^{\prime }+i\epsilon )}}\eqno(3.6)
$$%
one gets for the imaginary part, an analogous expression of 
\[
\displaylines{ {\rm{Im}}\  \Pi^{(\star,\star;2)}_R(Q)=\pi e^2m^4\int {{\rm d}^4P\over
(2\pi)^4}(1-2n_F(p_0))\sum_{s,s'=\pm 1}\ \biggl\lbrace -2\pi\beta_s(P)\beta_{s'}(P')W^{(2)}(P,P')\cr\hfill 
+\left(\alpha_s(P)\beta_{s'}(P')+\alpha_{s'}(P')\beta_s(P)\right)\left(-i\ {\rm
{disc}}_{p_0}W^{(2)}(P,P')\right)\biggr\rbrace\qquad(3.7)}
\]%
whose structure, the same as in the case of a single effective vertex
insertion, (3.4), allows some common and generic treatment of either cases :
\medskip In both (3.4) and (3.7), one has to cope with $\beta _{s}(P)$%
-distributions standing for the sum of a \textit{pole} part and a \textit{cut%
} part. Writing $\beta _{s}\equiv \beta _{s}^{(p)}+\beta _{s}^{(c)}$, the
textbook expressions are, \cite{[6]}, 
$$
-\beta _{s}^{(p)}(p_{0},p)=Z_{s}(p)\delta \left( p_{0}-\omega _{s}(p)\right)
+Z_{-s}(p)\delta \left( p_{0}+\omega _{-s}(p)\right) \eqno(3.8)
$$

$$
-\beta _{s}^{(c)}(p_{0},p)={\frac{m^{2}}{2p}}{\frac{(1-s{\frac{p_{0}}{p}}%
)\Theta (-P^{2})}{\left( p(1-s{\frac{p_{0}}{p}})-{\frac{m^{2}}{2p}}\left(
(1-s{\frac{p_{0}}{p}})\ln |{\frac{p_{0}+p}{p_{0}-p}}|+2s\right) \right) ^{2}+%
{\frac{\pi ^{2}m^{4}}{4p^{2}}}(1-s{\frac{p_{0}}{p}})^{2}}}\eqno(3.9) 
$$%
where for all $s=\pm 1$ and all $p$, the $Z_{s}(p)$ stand for the residues
at the quasi-particle poles. Now, in view of (3.8) and (3.9), three types of
contribution to (3.4) and (3.7) have to be considered :\newline
(i) Contributions involving the product of distributions $\beta
_{s}^{(p)}(P) $ and $\beta _{s^{\prime }}^{(p)}(P^{\prime })$, \newline
(ii) crossed contributions involving the product of distributions $\beta
_{s}^{(p)}(P)$ and $\beta _{s^{\prime }}^{(c)}(P^{\prime })$, \newline
(iii) contributions involving the product of distributions $\beta
_{s}^{(c)}(P)$ and $\beta _{s^{\prime }}^{(c)}(P^{\prime })$. \newline
\noindent In case (i), there is no infrared singularity problem at all,
because none of the quantities $P^{2}$, ${P^{\prime }}^{2}$, $2Q\!\cdot \!P$
and $p_{0}^{2}-2pxp_{0}+p^{2}$ can ever vanish over the whole integration
range. This will be exemplified to a large extent in the sequel. \smallskip
\noindent Case (iii) has been studied thoroughly and shown to lead to
singularity free contributions, \cite{[12]}.

\smallskip \noindent The intermediate, crossed case (ii) remains to be
studied. The two crossed possibilities contribute equally and the crossed
term $\beta _{s^{\prime }}^{(p)}(p_{0}^{\prime },p^{\prime })\times \beta
_{s}^{(c)}(p_{0},p)$ comes out to be proportional to the product of
distributions 
$$
\Theta (-P^{2})\times \{Z_{s^{\prime }}(p^{\prime })\delta \left(
p_{0}^{\prime }-\omega _{s^{\prime }}(p^{\prime })\right) +Z_{-s^{\prime
}}(p)\delta \left( p_{0}^{\prime }+\omega _{-s^{\prime }}(p^{\prime
})\right) \}\eqno(3.10) 
$$
where the residues at the quasi-particle poles read as
$$
Z_s(p)=\frac{\omega_s^2(p)-p^2}{2m^2}\eqno(3.11)
$$
The second delta is clearly incompatible with the overall $\Theta (-P^{2})$:
It would require that ${P^{\prime }}^{2}$ be strictly positive, whereas it
fixes a strictly negative term of $2Q\!\cdot \!P$. Since ${P^{\prime }}%
^{2}=P^{2}+2Q\!\cdot \!P$, this is impossible to satisfy at $P^{2}\leq 0$.
There is no incompatibility with the first delta function which fixes ${%
P^{\prime }}^{2}$ and $2Q\!\cdot \!P$ at strictly positive values, whereas ${%
P}^{2}$ can reach zero from below.

The constraint of $\delta (p_{0}^{\prime }-\omega _{s^{\prime
}}(p^{\prime }(x)))$ is common to both contributions appearing inside the
curly brackets of either (3.4) or (3.7). \smallskip Defining $p^{\prime }(x)=%
\sqrt{q^{2}+2pqx+p^{2}}$, with the two cosines $x={\widehat{q}}\!\cdot \!{%
\widehat{p}}$, and $y={\widehat{q}}\!\cdot \!{\widehat{p^{\prime }}}$, one
has $2Q\!\cdot \!P=2Q\!\cdot \!P^{\prime }=2q(\omega _{s^{\prime
}}(p^{\prime }(x))-yp^{\prime })$. Since $-1\leq y\leq 1$, and since $\omega
_{s^{\prime }}^{2}(p^{\prime })-p^{\prime }{}^{2}>0$,

\begin{figure}
 \caption{Dispersion relations : $
\protect\omega _{\pm }$/m as functions of k/m}
\label{fig:fig3}
 \begin{center}
   \includegraphics[width=\linewidth]{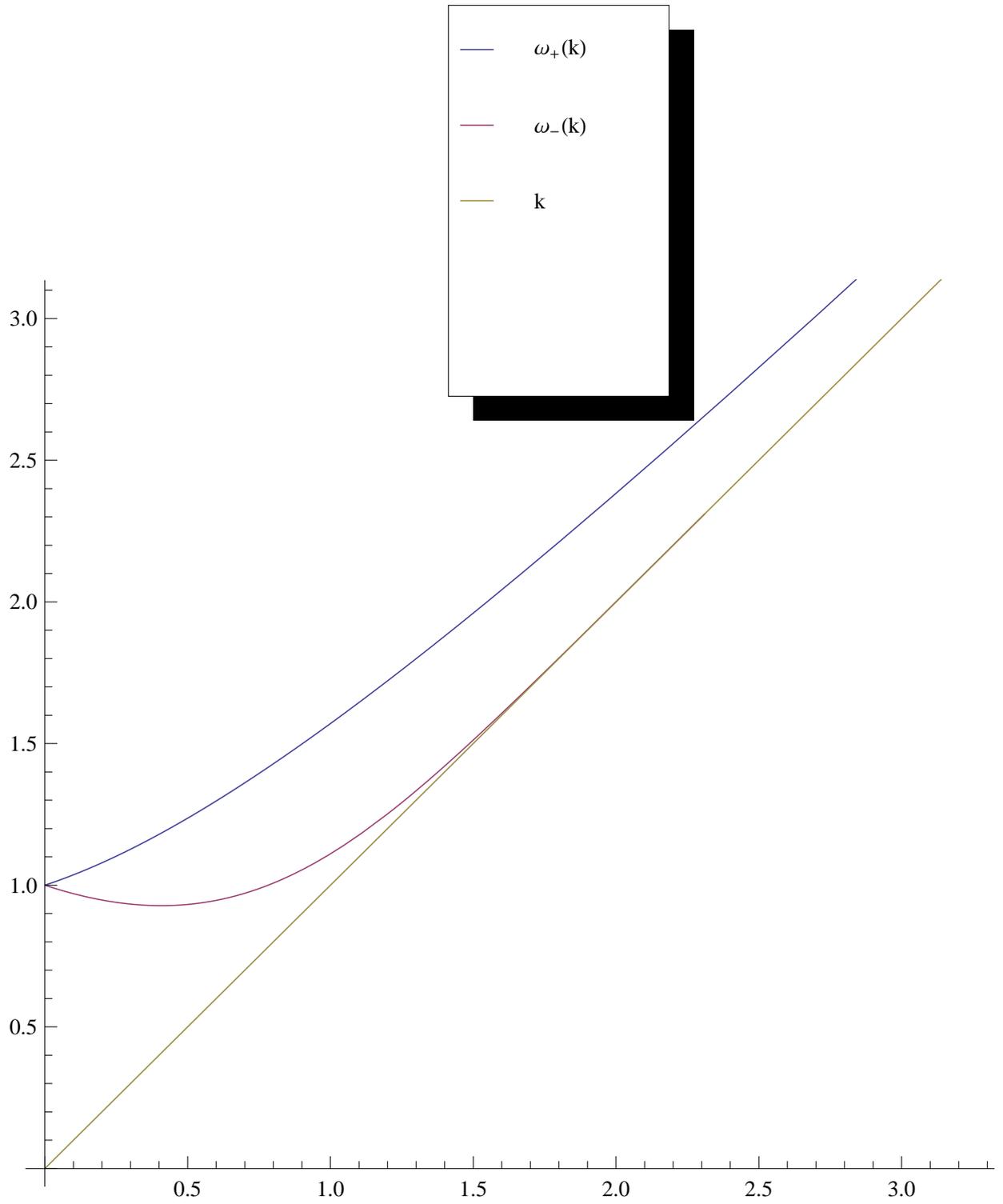}
 \end{center}
\end{figure}
one can deduce that 
$$
-1\leq x<{\frac{p_{0}}{p}}\eqno(3.12)
$$%
Then, since $P^{\prime }{}^{2}=\omega _{s^{\prime }}^{2}(p^{\prime
}(x))-p^{\prime }{}^{2}(x)>0$, so is therefore $P^{2}+2Q\cdot P$, which
gives 
$$
{\frac{p_{0}^{2}-p^{2}+2qp_{0}}{2qp}}>x\eqno(3.13)
$$%
and so 
$$
1+{\frac{p_{0}^{2}-p^{2}+2qp_{0}}{2qp}}={\frac{(p_{0}+p)(2q+p_{0}-p)}{2qp}}%
>1+x\geq 0\eqno(3.14)
$$%
The kinematics inherited from this common constraint restrict the
integration domain to the boundaries 
$$
{\mathcal{O}}(q)={\mathcal{O}}(p)=m\ ,\ \ \ p\leq q\ ,\ \ \ -p<p_{0}\ ,\ \ \
-1\leq x<{\frac{p_{0}}{p}}\eqno(3.15)
$$%
For the first terms in the curly brackets of both (3.4) and (3.7),
proportional to the products $\beta _{s^{\prime }}^{(p)}(p_{0}^{\prime
},p^{\prime })\times \beta _{s}^{(c)}(p_{0},p)$, an extra constraint of $%
\Theta (-P^{2})$ comes into play in view of (3.9), and modifies (3.15) into
an integration domain bounded by the relations 
$$
{\mathcal{O}}(q)={\mathcal{O}}(p)=m\ ,\ \ \ p\leq q\ ,\ \ \ -p<p_{0}\leq p\
,\ \ \ -1\leq x<{\frac{p_{0}}{p}}\eqno(3.16)
$$%
Actually, the arguments developed after (3.9) do not apply to the second
terms in the curly brackets of (3.4) and (3.7), the ones proportional to $%
\alpha _{s}(P)\times \beta _{s^{\prime }}^{(p)}(P^{\prime })$-contributions;
but it turns out that the terms of $\mathrm{disc}_{p_{0}}W^{(i)}(P,P^{\prime
})$, for $i\in \{1,2\}$, effectively restore the previous $\Theta (-P^{2})$%
-constraint, as can be read off (5.1.23) and (5.2.37) below, so as to make
of (3.16) the effective integration domain of the required resummation.

\smallskip Note that these inequalities automatically preclude any risk of
collinear singularity at $x=+1$, but not at $x=-1$, where the collinear
singularity was historically located \cite{[5]}.

\section{$1$- effective vertex calculations }

This case is given by Eq.(3.4), with the angular function $W^{(1)}$ given by
(3.3). In this case, the explicit calculation is quite simple. One gets 
\[
\displaylines{ W^{(1)}(P,P')={ss'\over pp'}+{s'\over p'}(1-s{p_0\over p}){1\over 2p}\ln{p_0+p\over p_0-p}+{s\over p}(1-s'{p'_0\over p'}){1\over 2p'}\ln{p'_0+p'\over p'_0-p'}\cr\hfill+(1-s{p_0\over
p})(1-s'{p'_0\over p'}){1\over 2Q\!\cdot\!P}\ln{{P'}^2\over P^2
}\qquad(4.1)} 
\]%
and so 
$$
\mathrm{{disc_{p_{0}}}}W^{(1)}(P,P^{\prime })=-i\pi \Theta (-P^{2})(1-s{%
\frac{p_{0}}{p}})\biggl\lbrace{\frac{s^{\prime }}{pp^{\prime }}}%
+(1-s^{\prime }{\frac{p_{0}^{\prime }}{p^{\prime }}}){\frac{\varepsilon
(p_{0})}{Q\!\cdot \!P}}\biggr\rbrace\eqno(4.2) 
$$

\smallskip \noindent The imaginary part of $\Pi _{R}^{(\star ,\star ;1)}(Q)$
can accordingly be written as the full expression 
\[
\displaylines{ {\rm{Im}}\  \Pi^{(\star,\star;1)}_R(Q)=-4\pi^2 e^2m^2\int {{\rm d}^4P\over
(2\pi)^4}(1-2n_F(p_0))\sum_{s,s'=\pm 1}\beta^{(c)}_s(P)\beta^{(p)}_{s'}(P')\cr\hfill\biggl\lbrace{ss'\over pp'}+{s'\over p'}(1-s{p_0\over p}){1\over 2p}\ln{p_0+p\over p_0-p}+{s\over p}(1-s'{p'_0\over p'}){1\over 2p'}\ln{p'_0+p'\over p'_0-p'}+(1-s{p_0\over
p})(1-s'{p'_0\over p'}){1\over 2Q\!\cdot\!P}\ln{{P'}^2\over P^2}\biggr\rbrace\cr\hfill 
-4\pi^2e^2m^2\int {{\rm d}^4P\over
(2\pi)^4}(1-2n_F(p_0))\Theta(-P^2)\cr\hfill \sum_{s,s'=\pm 1}\alpha_s(P)\beta^{(p)}_{s'}(P')(1-s{p_0\over p})\biggl\lbrace {s'\over pp'} +(1-s'{p'_0\over p'}){\varepsilon(p_0)\over Q\!\cdot\! P}\biggr\rbrace\qquad(4.3)}
\]%
and is to be integrated over the domain (3.16).

Let us begin with focusing on the first curly bracket of (4.3) :
Because $P^{\prime }{}^{2}$ as well as $2Q\!\cdot \!P$ are strictly
positive, only the logarithm of $p_{0}-p$ in the integrand, is able to yield
a diverging behavior, and there are two of them. Such a potentially
dangerous behavior is for example the one of 
\[
\displaylines{+2\pi e^2 \sum_{s,s'=\pm 1}\int {{\rm d}^3p\over
(2\pi)^3}\int_{-p}^{+p} {{\rm d}p_0\over
2\pi}(1-2n_F(p_0))\ {1\over 2Q\!\cdot\!P}\ln{(p_0-p)(p_0+p)\over P'^2}\cr\hfill \times\ \beta^{(c)}_s(p_0,p)(\omega_{s'}^2(p')-{p'}^2)\delta\left(p'_0-\omega_{s'}
(p')\right)(1-s{p_0\over
p})(1-s'{p'_0\over p'})\qquad(4.4)} 
\]%
where (3.11) has been used. However, at $s=+1$, the logarithmic divergence of the integrand
is suppressed by a factor of $1-p_{0}/p$ in $\beta _{+}^{(c)}(p_{0},p)$, so
that the case of $s=-1$ only must be considered whose potentially singular
part reads 
\[
\displaylines{+{2\pi e^2 }\int {p^2{\rm d}p\over
(2\pi)^2}\int_{-p}^{+p} {{\rm d}p_0\over
2\pi}(1-2n_F(p_0))\ (1+{p_0\over
p})\ \beta^{(c)}_-(p_0,p)\ \ln{p-p_0\over p}\cr\hfill \times\sum_{s'=\pm 1}\ \int_{-1}^{p_0\over p} {\rm{d}}x\ {\delta\left(q+p_0-\omega_{s'}
(p'(x))\right)\over 2q(\omega_{s'}(p'(x))-(q+px))}\ (\omega_{s'}^2(p'(x))-{p'^2(x)})(1-s'{\omega_{s'}
(p'(x))\over p'(x)})\qquad(4.5)} 
\]%
Because $\omega _{s^{\prime }}(p^{\prime }(x))$ is a fairly complicated,
implicit function of $x$, the last line of (4.5) is certainly hard to get
exactly. Fortunately this is not necessary either : It is sufficient that,
in a neighborhood of $p_{0}=p$, the second line of (4.5) defines a regular
function of $p_{0}$, say $F(p_{0})$. This condition is met indeed, and since
this situation is generic of all the potentially singular behaviors attached
to logarithms of $(p-p_{0})$, a proof is sketched in Appendix A.
\smallskip Then, in order to isolate the potentially singular behavior of
(4.5), one may re-write the second line of (4.5) as the sum $%
[F(p_{0})-F(p)]+F(p)$. Whereas the first term, $[F(p_{0})-F(p)]$ annihilates
the potentially divergent behavior of the logarithms, $[\ln (p-p_{0})/p]^{c}$%
, the second, $F(p)$, gives a contribution proportional to the would be
singular part of (4.5), that is to 
$$
\int^{p}\mathrm{d}p_{0}\ {\frac{\ln {\frac{p-p_{0}}{p}}}{\ln ^{2}{\frac{%
p-p_{0}}{p}}}}\sim \lim_{p_{0}=p}Li({\frac{p-p_{0}}{p}})=\lim_{p_{0}=p}{%
\frac{p-p_{0}}{p}}\int_{1}^{\infty }{\frac{1}{x^{2}}}\ {\frac{\mathrm{d}x}{%
\ln x+\ln {\frac{p}{p-p_{0}}}}}=0\eqno(4.6) 
$$%
where (3.9) has been used, and where $Li(x)$ is the \textit{%
Logarithm-integral function of x}, \cite{[13]}.

For the second term of (4.3), the one involving the discontinuity
in $p_{0}$, it is immediate to see that the same arguments apply, over the
same integration range (3.16), with the same conclusion.

Eventually, in contradistinction with the \textit{historical}
improper derivation, the imaginary part of $\Pi _{R}^{(\star ,\star ;1)}(Q)$
comes out singularity free when evaluated along the correct sequence of
discontinuity and angular average operations.

\section{$2$-effective vertex calculations }

Though crucial, since the original collinear singularity is explicitly due
to it, this case is far more difficult because, as an unavoidable step, the
angular function $W^{(2)}(P,P^{\prime })$ of (3.6) must be known exactly.
Let us begin with recalling this function 
$$
W^{(2)}(P,P^{\prime })=\int {\frac{\mathrm{d}{\widehat{K}}}{4\pi }}\int {%
\frac{\mathrm{d}{\widehat{K}^{\prime }}}{4\pi }}\ {\widehat{K}}\!\cdot \!{%
\widehat{K}^{\prime }}{\frac{{\widehat{K}}\!\cdot \!{\widehat{P}}_{s}\ {%
\widehat{K}^{\prime }}\!\cdot \!{\widehat{P}^{\prime }}_{s^{\prime }}+{%
\widehat{K}}\!\cdot \!{\widehat{P}^{\prime }}_{s^{\prime }}\ {\widehat{K}%
^{\prime }}\!\cdot \!{\widehat{P}}_{s}-{\widehat{K}}\!\cdot \!{\widehat{K}%
^{\prime }}{\widehat{P}}_{s}\!\cdot \!{\widehat{P}^{\prime }}_{s^{\prime }}}{%
({\widehat{K}}\!\cdot \!P+i\epsilon )({\widehat{K}}\!\cdot \!P^{\prime
}+i\epsilon )({\widehat{K}^{\prime }}\!\cdot \!P+i\epsilon )({\widehat{K}%
^{\prime }}\!\cdot \!P^{\prime }+i\epsilon )}}\eqno(3.6) 
$$%
and define $W^{(2)}(P,P^{\prime })=W_{1}^{(2)}(P,P^{\prime
})+W_{2}^{(2)}(P,P^{\prime })$. The function $W_{1}^{(2)}(P,P^{\prime })$
corresponds to the first two terms in the numerator of (3.6). They are
symmetric in the exchange of $P$ and $P^{\prime }$ and contribute equally.
That is, 
\[
\displaylines{W^{(2)}_1(P,P')=2\int_{{\widehat K}}\int_{{\widehat K'}}\
{{\widehat K}\!\cdot\!{\widehat K'} \over ({\widehat K}\!\cdot\!
P'+i\epsilon)({\widehat K'}\!\cdot\! P+i\epsilon)}\ \biggl\lbrace {ss'\over
pp'}+{s'\over p'}(1-s{p_0\over p}){1\over {\widehat K}\!\cdot\!
P+i\epsilon}\cr\hfill +{s\over p}(1-s'{p'_0\over p'}){1\over {\widehat
K'}\!\cdot\! P'+i\epsilon}+(1-s{p_0\over p})(1-s'{p'_0\over p'}){1\over
({\widehat K}\!\cdot\! P+i\epsilon)({\widehat K'}\!\cdot\!
P'+i\epsilon)}\biggr\rbrace \qquad(5.1)} 
\]%
whereas $W_{2}^{(2)}(P,P^{\prime })$ is the function 
$$
W_{2}^{(2)}(P,P^{\prime })=-{\widehat{P}}_{s}\!\cdot \!{\widehat{P}^{\prime }%
}_{s^{\prime }}\int_{{\widehat{K}}}\int_{{\widehat{K}^{\prime }}}\ {\frac{({%
\widehat{K}}\!\cdot \!{\widehat{K}^{\prime }})^{2}}{({\widehat{K}}\!\cdot
\!P+i\epsilon )({\widehat{K}}\!\cdot \!P^{\prime }+i\epsilon )({\widehat{K}%
^{\prime }}\!\cdot \!P+i\epsilon )({\widehat{K}^{\prime }}\!\cdot
\!P^{\prime }+i\epsilon )}}\eqno(5.2) 
$$%
We now cope exclusively with $W_{1}^{(2)}(P,P^{\prime })$. The calculation
of $W_{2}^{(2)}(P,P^{\prime })$ being \textquotedblleft orders of magnitude"
more difficult will be dealt with in subsection 5.2.

\subsection{{The case of $W_{1}^{(2)}(P,P^{\prime })$}}

The contribution of $W_{1}^{(2)}(P,P^{\prime })$ to $\mathrm{Im}\ \Pi
_{R}^{(\star ,\star ;2)}(Q)$, is obtained by substituting $%
W_{1}^{(2)}(P,P^{\prime })$ for $W^{(2)}(P,P^{\prime })$ in (3.7), and we
begin with the $\beta _{s^{\prime }}^{(p)}(P^{\prime })\times \beta
_{s}^{(c)}(P)$ -term.

\smallskip - From (5.1), a first part, coming from the term ${ss^{\prime}/
pp^{\prime}}$ contributes to (3.7) the amount 
\[
\displaylines{-2\pi^2 e^2\int_P(1-2n_F(p_0))\sum_{s,s'=\pm 1}\ \beta^{(c)}_{s}(p_0,p)  \  {\omega_{s'}^2(p')-{p'}^2\over 2m^2}\delta\left(p'_0-\omega_{s'}
(p')\right)\cr\hfill \times {ss'\over pp'}\ \Sigma_R(P)\cdot \Sigma_R(P')\qquad(5.1.3)}
\]%
where the "self energy four-vector" has components, 
$$
\Sigma_\alpha^0(P)={\frac{m^2}{p}}Q_0({\frac{p_0}{p}})\ ,\ \ \ \
\Sigma_\alpha^i(P)=({\frac{{\vec p}^i}{p}}\equiv {\widehat p}^i) {\frac{m^2}{%
p}}Q_1({\frac{p_0}{p}})\eqno(5.1.4) 
$$%
with $Q_0$ and $Q_1$, the Legendre functions 
$$
Q_1(x)=xQ_0(x)-1\ ,\ \ \ \ \ \ Q_0(x)={\frac{1}{2}}\ln{\frac{x+1}{x-1}}\eqno%
(5.1.5) 
$$%
The label $\alpha=\{R,A\}$ keeps on denoting one of the two \textit{Retarded}
or \textit{Advanced} specifications of the real time formalism being used,
and in the right hand sides of (5.1.4) these specifications are encoded in
the logarithmic determinations. Because of the delta distribution, $%
\delta\left(p^{\prime}_0-\omega_{s^{\prime}} (p^{\prime})\right)$, one of
the self energies of (5.1.3) is the regular function, $\Sigma_R(\omega_{s^{%
\prime}}(p^{\prime}), p^{\prime})$, whereas the other one, $\Sigma_R(p_0,p)$%
, entails the logarithmic components of (5.1.4) and (5.1.5). \smallskip Over
the integration range of $-p< p_0\leq +p$, though themselves divergent, but
logarithmically only, these components lead to the same singularity free
result as obtained in Section 4, Eq.(4.6).

\bigskip - For the second term in the big parenthesis of (5.1) one can take
advantage of Eq.(4.14) of Ref.12, to find 
\[
\displaylines{{s'\over p'}(1-s{p_0\over p})\int_{{\widehat K}}\int_{{\widehat K'}}\ {{\widehat K}\!\cdot\!{\widehat K'} \over ({\widehat K}\!\cdot\! P'+i\epsilon)({\widehat K'}\!\cdot\! P+i\epsilon)}\ {1\over {\widehat K}\!\cdot\! P+i\epsilon}={s'\over p'}(1-s{p_0\over p})\biggl\lbrace {1\over p^2}Q_1({p_0\over p}){1\over 2p'}\ln{p'_0+p'\over p'_0-p'}\cr\hfill
+{1\over p}\left({p_0\over p}-{P^2\over p^2}Q_0({p_0\over p})\right){1\over 2Q\!\cdot\!P}\ln{{P'}^2\over P^2}\biggr\rbrace\qquad(5.1.6)}
\]%
so that, when plugged back into (3.7), one gets 
\[
\displaylines{-2\pi^2 e^2m^4\int  {{\rm d}^3p\over
(2\pi)^3}\int_{-p}^{+p}  {{\rm d}p_0\over
2\pi}(1-2n_F(p_0))\sum_{s,s'=\pm 1}\ \beta^{(c)}_{s}(p_0,p)  \  {\omega_{s'}^2(p')-{p'}^2\over 2m^2}\delta\left(p'_0-\omega_{s'}
(p')\right)\cr\hfill \times{s'\over p'}(1-s{p_0\over p})\biggl\lbrace {1\over p^2}Q_1({p_0\over p}){1\over 2p'}\ln{\omega_{s'}
(p')+p'\over \omega_{s'}
(p')-p'}\cr\hfill
+{1\over p}\left({p_0\over p}-{P^2\over p^2}Q_0({p_0\over p})\right){1\over 2q(\omega_{s'}(p')-p'y)}\ln{\omega_{s'}^2(p')-{p'}^2\over P^2}\biggr\rbrace \qquad(5.1.7)}
\]%
\medskip \noindent where we have used $2Q\!\cdot \!P=2Q\!\cdot \!P^{\prime
}=2q(\omega _{s^{\prime }}(p^{\prime })-p^{\prime }y)>0$, in order to
emphasize the non vanishing character of this factor. Again, the integrand
\textquotedblleft wildest behavior\textquotedblright\ is the one of the
logarithms of $p_{0}-p$, which, integrated over the interval $-p<p_{0}\leq
+p $ leads to regular contributions.

\bigskip - The same conclusion holds for the third term in the big
parenthesis of (5.1), which, easily obtained out of the second one, is
quoted here for the sake of completeness, 
\[
\displaylines{-2\pi^2 e^2m^4\int  {{\rm d}^3p\over
(2\pi)^3}\int_{-p}^{+p}  {{\rm d}p_0\over
2\pi}(1-2n_F(p_0))\sum_{s,s'=\pm 1}\ \beta^{(c)}_{s}(p_0,p)  \  {\omega_{s'}^2(p')-{p'}^2\over 2m^2}\delta\left(p'_0-\omega_{s'}
(p')\right)\cr\hfill \times{s\over p}(1-s'{p'_0\over p'})\biggl\lbrace {1\over {p'}^2}Q_1({\omega_{s'}
(p')\over p'}){1\over 2p}\ln{p_0+p\over p_0-p}\cr\hfill
+{1\over p'}\left({\omega_{s'}
(p')\over p}-{{P'}^2\over {p'}^2}Q_0({\omega_{s'}
(p')\over p'})\right){1\over 2q(\omega_{s'}(p')-p'y)}\ln{\omega_{s'}^2(p')-{p'}^2\over P^2}\biggr\rbrace \qquad(5.1.8)}
\]%
Note that the last term of (5.1.8) is not induced by an error of \textit{%
copy and paste}, but reflects the symmetry of $1/2Q\!\cdot \!P\times \ln {%
P^{\prime }}^{2}\!/\!P^{2}$ under the exchange of $P^{\prime }$ and $P$,
since $2Q\!\cdot \!P={P^{\prime }}^{2}-P^{2}$.

\bigskip - With the fourth term in the big parenthesis of (5.1), things
become more involved. This term in effect, entails the following angular
integration 
$$
(1-s{\frac{p_{0}}{p}})(1-s^{\prime }{\frac{p_{0}^{\prime }}{p^{\prime }}}%
)\int_{{\widehat{K}}}\int_{{\widehat{K}^{\prime }}}\ {\frac{{\widehat{K}}%
\!\cdot \!{\widehat{K}^{\prime }}}{({\widehat{K}}\!\cdot \!P+i\epsilon )({%
\widehat{K}}\!\cdot \!P^{\prime }+i\epsilon )({\widehat{K}^{\prime }}\!\cdot
\!P+i\epsilon )({\widehat{K}^{\prime }}\!\cdot \!P^{\prime }+i\epsilon )}}%
\eqno(5.1.9) 
$$%
One can take advantage of the calculations of Ref.12, in particular of the
angular identity ($R=(r_{0},{\vec{r}})$, $r=|{\vec{r}}|$) 
$$
\int {\frac{\mathrm{d}{\widehat{K}}}{4\pi }}\ {\frac{{\widehat{K}}^{i}}{({%
\widehat{K}}\!\cdot \!R+i\epsilon )^{2}}}={\frac{r^{i}}{r^{2}}}\left( {\frac{%
1}{2r}}\ln {\frac{r_{0}+r}{r_{0}-r}}-{\frac{r_{0}}{R^{2}+i\epsilon r_{0}}}%
\right) \eqno(5.1.10) 
$$%
an euclidean version of which can be found in \cite{[6]}. Using it, the
result can be cast into the form 
\[
\displaylines{\int{{\rm d}{\widehat K}\over
4\pi}{{\widehat
K}^\mu\over ({\widehat K}\!\cdot\! P+i\epsilon)({\widehat
K}\!\cdot\! P'+i\epsilon)}\ \int{{\rm d}{\widehat K'}\over 4\pi}{{\widehat
K'}_\mu\over ({\widehat K'}\!\cdot\! P+i\epsilon)({\widehat
K'}\!\cdot\! P'+i\epsilon)}=\cr\hfill -\sum_{i,j=0}^3 \ \left(\sum_{k=-2}^{+1} a^k_{ij}\ (2Q\!\cdot\!
P)^k\right)
F_iF_j\qquad(5.1.11)} 
\]%
where the following definitions are used : 
$$
r^{2}(s)=p^{2}+2pqxs+q^{2}s^{2}\ ,\ \ \ \ R^{2}(s)=P^{2}+Zs\ ,\ \ \ \
Z=2Q\!\cdot \!P\eqno(5.1.12) 
$$%
and where the $F_{i}^{\prime }s$ stand for the four functions

$$
F_{0}(P,Q)=\int_{0}^{1}{\frac{\mathrm{d}s}{R^{2}(s)}}={\frac{1}{2Q\!\cdot \!P%
}}\ln {\frac{P^{\prime }{}^{2}}{P^{2}}}\eqno(5.1.13) 
$$%
$$
F_{2}(P,Q)=\int_{0}^{1}{\frac{\mathrm{d}s}{r^{2}(s)}}={\frac{1}{qp\sqrt{%
1-x^{2}}}}\arctan {\frac{q\sqrt{1-x^{2}}}{p+qx}}\eqno(5.1.14) 
$$%
$$
F_{1}-{\frac{px}{q}}F_{2}=\int_{0}^{1}{\frac{s\mathrm{d}s}{r^{2}(s)}}={\frac{%
1}{2q^{2}}}\ln {\frac{p^{\prime }{}^{2}}{p^{2}}}-{\frac{px}{q}}F_{2}\eqno%
(5.1.15) 
$$%
\[
\displaylines{F_3(P,Q)=\int_0^1{{\rm{d}}s\over r^2(s) R^2(s)}={1\over {
q^2}\ {
\left({p_0^2-2pxp_0+p^2}\right)^2}}\biggl\lbrace(q^2P^2-pqxZ)F_2\cr\hfill +
Z^2F_0- q^2ZF_1\biggr\rbrace\qquad (5.1.16)} 
\]%
Eventually, the non vanishing $a_{ij}^{k}$-coefficients of (5.1.11) are
polynomials in $p_{0}$ 
$$
a_{22}^{-2}=-q^{2}P^{2},\ a_{22}^{-1}=qp_{0}\eqno(5.1.17) 
$$%
$$
a_{33}^{-2}=-q^{2}(P^{2})^{3},\ a_{33}^{-1}={\frac{5}{2}}qp_{0}(P^{2})^{2},\
a_{33}^{0}=-{\frac{9}{4}}(P^{2})^{2}-{\frac{5}{2}}p^{2}P^{2},\ a_{33}^{1}={%
\frac{p_{0}(3P^{2}+4p^{2})}{4q}}\eqno(5.1.18) 
$$%
$$
a_{02}^{0}=1\eqno(5.1.19) 
$$%
$$
a_{03}^{-1}=-qp_{0}P^{2},\ a_{03}^{0}={\frac{3}{2}}P^{2},\ a_{03}^{1}=-{%
\frac{p_{0}}{q}}\eqno(5.1.20) 
$$%
$$
a_{23}^{-2}=2q^{2}(P^{2})^{2},\ a_{23}^{-1}=-4qp_{0}P^{2},\ a_{23}^{0}={%
\frac{11}{4}}P^{2}+{\frac{3}{2}}p^{2},\ a_{23}^{1}=-{\frac{p_{0}}{q}}\eqno%
(5.1.21) 
$$

\medskip \noindent Over the integration domain (3.16), since $Z=2Q\!\cdot
\!P $ does not vanish, the potentially singular behaviors of (3.7) are to be
looked for in relation to the behaviors of the $F_{i}^{\prime }s$.

\medskip - The case of $(F_{2})^{2}$-contributions, with associated
coefficients (5.1.17), is dealt with easily. Since $F_{2}$ is a perfectly
regular function of its variables, $(F_{2})^{2}$-contributions to (3.7) are
singularity free.

\medskip - And so is, in the same vein, the $F_{0}F_{2}$-contribution to
(3.7), corresponding to the coefficient (5.1.19).

\medskip - For the function $F_{3}$, one has a denominator of $({%
p_{0}^{2}-2pxp_{0}+p^{2}})^{2}$ which has no zeros in the integration range
(3.16). The potentially singular most behavior of $F_{3}$ is again the one
of $F_{0}$, with its $\ln [(p_{0}-p)/p]$-term. It results that not only
contributions to (3.7) of type $(F_{2}F_{3})$, with associated coefficients
(5.1.21), but also $(F_{3})^{2}$- and $(F_{0}F_{3})$-contributions to (3.7),
respectively associated to coefficients (5.1.18) and (5.1.20), are
singularity-free.

\bigskip The $W_{1}^{(2)}(P,P^{\prime })$ -contributions to the 2-effective
vertex part of the soft photon emission rate involve another piece, the one
associated to the term $-i\mathrm{disc}_{p_{0}}W_{1}^{(2)}(P,P^{\prime })$.

\smallskip

As made clear by a simple inspection of the $p_{0}$-dependences in (5.1.3),
(5.1.7) and (5.1.8), taking the discontinuity in $p_{0}$ just amounts to
substitute a term of $\pm i\pi \Theta (-P^{2})$ for a logarithmic term of $%
\ln [(p_{0}-p)/p]$, all of the other discontinuities being zero or giving
zero: Such is for example the case of the discontinuity proportional to $%
\delta (2Q\!\cdot \!P)$ which has no support in (3.16).

\smallskip - No singular contributions are therefore generated by (5.1.3),
(5.1.7) and (5.1.8), when the discontinuity in $p_{0}$ is taken.

\smallskip - The last and more complicated term involves the discontinuity
in $p_{0}$ of (5.1.9), that is 
$$
-i\mathrm{disc}_{p_{0}}(5.1.9)=+i(1-s{\frac{p_{0}}{p}})(1-s^{\prime }{\frac{%
p_{0}^{\prime }}{p^{\prime }}})\ \mathrm{disc}_{p_{0}}\sum_{i,j=0}^{3}\
\left( \sum_{k=-2}^{+1}a_{ij}^{k}\ (2Q\!\cdot \!P)^{k}\right) F_{i}F_{j}\eqno%
(5.1.22) 
$$%
Now this is simple also, because the $a_{ij}^{k}$ of (5.1.17)-(5.1.21) are
polynomials in $p_{0}$, and because $\delta (2Q\!\cdot \!P)$ has no support
in the integration domain. Moreover, one has $\mathrm{disc}_{p_{0}}F_{1}=%
\mathrm{disc}_{p_{0}}F_{2}=0$, whereas $\mathrm{disc}_{p_{0}}F_{0}$ is
restricted to $=\pm i\pi \Theta (-P^{2})/Z$ because, as stated above, $%
\delta (2Q\!\cdot \!P)$ has no support. The discontinuity of $F_{3}$ is
restricted to $\pm i\pi Z\Theta (-P^{2})/(p_{0}^{2}-2pxp_{0}+p^{2})^{2}$
because, as demonstrated below, Eq.(5.2.38), $\delta
(p_{0}^{2}-2pxp_{0}+p^{2})$ and $\delta (p_{0}^{\prime }-\omega _{s^{\prime
}}(p^{\prime }))$ are incompatible constraints. One gets eventually 
\[
\displaylines{ -i{\rm{disc}}_{p_0}(5.1.9)=\mp (1-s{p_0\over p})(1-s'{p'_0\over p'})\biggl\lbrace (a_{02}^0)\ {\pi\Theta(-P^2)\over Z}F_2\cr\hfill
+(\ a_{03}^{-1}Z^{-1}+a_{03}^0+a_{03}^{1}Z^{1}\ )\ (\ {\pi\Theta(-P^2)\over Z}F_3+{\pi\Theta(-P^2)\over (p_0^2-2pxp_0+p^2)^2}\ln{P'^2\over P^2}\ )\cr\hfill
+(\ a_{23}^{-2}Z^{-2}+a_{23}^{-1}Z^{-1}+a_{23}^0+a_{23}^{1}Z^{1}\ )\ {\pi\Theta(-P^2)ZF_2\over (p_0^2-2pxp_0+p^2)^2}\cr\hfill
+2(\ a_{33}^{-2}Z^{-2}+a_{33}^{-1}Z^{-1}+a_{33}^0+a_{33}^{1}Z^{1}\ )\ {\pi\Theta(-P^2)ZF_3\over (p_0^2-2pxp_0+p^2)^2}\biggr\rbrace\qquad(5.1.23)}
\]%
For the same reasons as before, it should be clear that when plugged back
into (3.7), these terms, over (3.16), do not induce any singular behavior of the
subsequent integrations on $x$, $p_{0}$ and $p$.

\subsection{{The case of $W_{2}^{(2)}(P,P^{\prime })$} }

We now come to the last and most tedious angular integration, the one
defining the function $W_{2}^{(2)}(P,P^{\prime })$ of (5.2). Writing it as 
$$
W_{2}^{(2)}(P,P^{\prime })=\int_{0}^{1}\mathrm{d}s\int_{0}^{1}\mathrm{d}%
s^{\prime }\int {\frac{\mathrm{d}{\widehat{K}}}{4\pi }}\int {\frac{\mathrm{d}%
{\widehat{K}^{\prime }}}{4\pi }}{\frac{1-2{\widehat{K}}^{i}{\widehat{K}%
^{\prime }}_{i}+{\widehat{K}}^{i}{\widehat{K}}^{j}{\widehat{K}^{\prime }}_{i}%
{\widehat{K}^{\prime }}_{j}}{({\widehat{K}}\!\cdot \!R(s)+i\epsilon )^{2}({%
\widehat{K}^{\prime }}\!\cdot \!R(s^{\prime })+i\epsilon )^{2}}}\eqno(5.2.1) 
$$%
it is possible to add and subtract a $+1$ in the numerator of (5.2.1), to
get 
\[
\displaylines{W^{(2)}_2(P,P')=2\int_{{\widehat K}}\int_{{\widehat K'}}\ {{\widehat K}\!\cdot\!{\widehat K'} \over ({\widehat K}\!\cdot\! P+i\epsilon)({\widehat K}\!\cdot\! P'+i\epsilon)({\widehat K'}\!\cdot\! P+i\epsilon)({\widehat K'}\!\cdot\! P'+i\epsilon)}-{1\over Z^2}\ln^2{P'^2\over P^2}\cr\hfill +\int_0^1{\rm{d}}s\int_0^1{\rm{d}}s'\int{{\rm d}{\widehat K}\over
4\pi}\int{{\rm d}{\widehat K'}\over
4\pi}{{\widehat K}^i{\widehat K}^j{\widehat K'}_i{\widehat K'}_j\over ({\widehat
K}\!\cdot\!R(s)+i\epsilon)^2({\widehat K'}\!\cdot\!R(s')+i\epsilon)^2}\qquad(5.2.2)}
\]%
In the first line, the double angular integral is the one appearing already
in (5.1.9), which as we have just seen, causes no singularity problem, and so is also the case of
the second term. One can accordingly focus on the new, third term in the
second line of (5.2.2).

\bigskip \noindent This new term can be dealt with the help of the angular
identity \cite{[12]} ($R(s)=P+sQ$)

$$
\int {\frac{\mathrm{d}{\widehat{K}}}{4\pi }}\ {\frac{{\widehat{K}}^{i}{%
\widehat{K}}^{j}}{({\widehat{K}}\!\cdot \!R(s)+i\epsilon )^{2}}}=-{\frac{%
g^{ij}}{r^{2}}}Q_{1}({\frac{r_{0}}{r}})-{\frac{r^{i}r^{j}}{r^{2}}}\left( {%
\frac{3}{r^{2}}}Q_{1}({\frac{r_{0}}{r}})-{\frac{1}{R^{2}(s)+i\epsilon r_{0}}}%
\right) \eqno(5.2.3) 
$$%
an euclidean version of which can be found in \cite{[6]}. When using that
identity, one finds for the third term of (5.2.2), a sum of five fairly
complicated contributions 
\[
\displaylines{W^{(2)}_2(P,P')\ \ni\  -3\left( \int_0^1{\rm{d}}s\ {Q_1(R(s))\over
r^2(s)}\right)^2+{{2}}\left( \int_0^1{\rm{d}}s\ {Q_1(R(s))\over
r^2(s)}\right)\left( \int_0^1\ {{\rm{d}}s'\over
R^2(s')+i\epsilon r_0(s')}\right)\cr\hfill +9\int_0^1{\rm{d}}s\ {Q_1(R(s))\over
r^2(s)}\int_0^1{\rm{d}}s'\ [{\widehat r(s)}\!\cdot\!{\widehat r(s')}]^2{Q_1(R(s'))\over
r^2(s')}\ -6\int_0^1{\rm{d}}s\ {Q_1(R(s))\over
r^2(s)}\int_0^1{\rm{d}}s'\ {[{\widehat r(s)}\!\cdot\!{\widehat r(s')}]^2\over
R^2(s')+i\epsilon r_0(s')}\cr\hfill +\int_0^1{{\rm{d}}s\over (R^2(s)+i\epsilon r_0(s))}\int_0^1{\rm{d}}s'\
{[{\widehat r(s)}\!\cdot\!{\widehat r(s')}]^2\over (R^2(s')+i\epsilon r_0(s'))}\qquad(5.2.4)}
\]%
\medskip \noindent In order to express any of the five terms composing
(5.2.4), and besides the definitions (5.1.12)-(5.1.16), the following
integrals are needed : 
\[
\displaylines{I_3=\int_0^1{\rm{d}}s\ {\ln X(s)\over r^3(s)}={1\over
qp^2(1-x^2)}\left((q+px){\ln X'\over
p'}-px{\ln X\over p}\right)\cr\hfill -{ZF_0\over qp^2(1-x^2)}+2p_0F_3+{2q}{F_2-P^2F_3\over Z}\qquad(5.2.5)}
\]

\[
\displaylines{{ I'_3}=\int_0^1s{\rm{d}}s\ {\ln X(s)\over r^3(s)}={1\over
pq^2(1-x^2)}\left(p{\ln X\over  p}-(p+qx){\ln X'\over p'}\right)\cr\hfill +{x\over
pq^2(1-x^2)}ZF_0-2{p^2\over q}F_3+{2}(p_0-2px){F_2-P^2F_3\over Z}\qquad(5.2.6)}
\]
where 
$$
X(s)={\frac{r_0(s)+r(s)}{r_0(s)-r(s)}}\ ,\ \ \ X=X(0)={\frac{p_0+p}{p_0-p}}\
,\ \ \ X^{\prime}=X(1)={\frac{p^{\prime}_0+p^{\prime}}{p^{\prime}_0-p^{%
\prime}}}\eqno(5.2.7) 
$$
\medskip\noindent Then, it is possible to give the final expression for the
second term of (5.2.4). It is 
$$
2F_0\int_0^1\mathrm{d}s\ {\frac{Q_1(R(s))}{r^2(s)}}={\frac{F_0}{pq(1-x^2)}}%
\left((p_0x-p+{\frac{Z}{2p}}){\frac{\ln X^{\prime}}{p^{\prime}}}-(p_0x-p){%
\frac{\ln X}{p}}\right)-{\frac{1}{2}}{\frac{ (ZF_0)^2}{p^2q^2(1-x^2)}}\eqno%
(5.2.8) 
$$
and also, for the first term of (5.2.4) : 
\[
\displaylines{\!\!\!\!\!\!\!\!\!\!\!\!\!\!\!\!\!\!\!\!\! \!\!\!\!\!\!\!\!\!\!\!\!\!\!\!\!\!\!\!\!\!
\!\!\!\!\!\!\!\!\!\!\!\!\!\!\!\!\!\!\!\!\! \!\!\!\!\!\!\!\!\!\!\!\!\!\!\!\!\!\!\!\!\!
\!\!\!\!\!\!\!\!\!\!\!\!\!\!\!\!\!\!\!\!\! \!\!\!\!\!\!\!\!\!\!\!\!\!\!\!\!\!\!\!\!\! -3\left( \int_0^1{\rm{d}}s\
{Q_1(R(s))\over
r^2(s)}\right)^2\cr\hfill ={-3\over 4p^2q^2(1-x^2)^2}\left(-{1\over 2}{Z^2F_0\over pq}+(p_0x-p+{Z\over 2p}){\ln
X'\over p'}-(p_0x-p){\ln X\over p}\right)^2\qquad(5.2.9)} 
\]
\bigskip - The fifth term of (5.2.4) can be cast into the form 
$$
\sum_{i,j=0}^3 \ \left(\sum_{k=-2}^{+1} b^k_{ij}\ Z^k\right) F_iF_j\eqno%
(5.2.10) 
$$%
where the non vanishing $b^k_{ij}$ are given by the array 
$$
b^0_{00}=1\eqno(5.2.11) 
$$
$$
b^{-2}_{22}=2{q^2p^2(1-x^2)}\eqno(5.2.12) 
$$
$$
b^{-2}_{33}=2{q^2p^2(1-x^2)(P^2)^2}\ ,\ \ \ b_{33}^{-1}=-4qp^3x(1-x^2)P^2\
,\ \ \ b_{33}^0=2p^4(1-x^2) \eqno(5.2.13) 
$$
$$
b_{03}^0=-2p^2(1-x^2)\eqno(5.2.14) 
$$
$$
b_{23}^{-2}=-4q^2p^2(1-x^2)P^2\ ,\ \ \ b_{23}^{-1}=4qp^3x(1-x^2)\eqno%
(5.2.15) 
$$
\bigskip\bigskip - The fourth term of (5.2.4) reads 
\[
\displaylines{-6F_0\biggl\lbrace -{1\over 2}F_2+{1\over 2q}({q+px\over p'^2}-{x\over p}) + {p_0\over 2}I_3 + {q\over 2}I'_3
-{p_0p^2(1-x^2)\over 2}I_5 -{qp^2(1-x^2)\over 2}I'_5\biggr\rbrace \cr\hfill
-6p^2(1-x^2)F_3\biggl\lbrace  - {1\over p'^2} + (px-{p_0\over 2})I_3 - {q\over 2}I'_3 +
{p^2\over 2q}ZI_5 + pq\left(p({{1}}-x^2)+{x\over 2q}Z\right)I'_5\biggr\rbrace \cr\hfill
-12qp^2(1-x^2){F_2-P^2F_3\over Z}\biggl\lbrace {1\over 2qp}({p\over p'^2}-{1\over p}) + {1\over 2}I_3 + {p(p_0x-p)\over 2}I_5 + {Z\over 4}I'_5\biggr\rbrace\qquad(5.2.16)}
\]
where two extra more complicated integrals are needed : 
$$
I_5=\int_0^1\mathrm{d}s\ {\frac{\ln X(s)}{r^5(s)}}\ ,\ \ \
I^{\prime}_5=\int_0^1s\mathrm{d}s\ {\frac{\ln X(s)}{r^5(s)}}\eqno(5.2.17) 
$$
One finds 
\[
\displaylines{I_5={1\over
3qp^2(1-x^2)}\left({q+px\over
p'^3}\ln X'-{x\over p^2}\ln X\right)+{2\over
3qp^4(1-x^2)^2}\left({q+px\over
p'}\ln X'-{x}\ln X\right)\cr\hfill
+{4q\over 3p^2(1-x^2)}{F_2-P^2F_3\over Z}+{2(p_0+px)\over 3p^2(1-x^2)}{F_3} -{2ZF_0\over 3qp^4(1-x^2)^2}\cr\hfill
+{2q\over 3Z}\left(\int_0^1{ds\over r^4(s)}+(p_0^2-2p_0px+p^2)\int_0^1{ds\over R^2 r^4(s)}\right)\qquad(5.2.18)}
\]
\[
\displaylines{I'_5={-1\over
3q^2p(1-x^2)}\left({p+qx\over
p'^3}\ln X'-{1\over p^2}\ln X\right)-{2x\over
3q^2p^3(1-x^2)^2}\left({q+px\over
p'}\ln X'-{x}\ln X\right)\cr\hfill
+{2x\over 3q^2p^3(1-x^2)^2}{ZF_0 }-{2x(p_0+px)\over 3qp(1-x^2)}F_3-{4x\over 3p(1-x^2)}{F_2-P^2F_3\over Z}\cr\hfill
+({1\over 3q}-{2qpx\over 3Z})\int_0^1{ds\over r^4(s)}\cr\hfill
-2p_0{p_0^2-2p_0px+p^2\over 3Z}\int_0^1{ds\over R^2(s) r^4(s)}\qquad(5.2.19)}
\]
with 
$$
\int_0^1 {\frac{ds}{r^4(s)}}={\frac{1}{2qp^2(1-x^2)}}\left({\frac{px+q}{%
{p^\prime}^2}}-{\frac{x}{p}} +qF_2\right)\eqno(5.2.20) 
$$
$$
\int_0^1 {\frac{sds}{r^4(s)}}={\frac{1}{2pq^2(1-x^2)}}\left( {\frac{1}{p}}-{%
\frac{p+qx}{{p^\prime}^2}}- qxF_2\right)\eqno(5.2.21) 
$$%
and 
\[
\displaylines{\int_0^1 {ds\over R^2(s)r^4(s)}={1\over q^2(p_0^2-2p_0px+p^2)^2}\biggl\lbrace Z^2F_3+q^2\left(P^2-{2px\over q}Z\right)\int_0^1 {ds\over r^4(s)}\cr\hfill
-q^2Z\int_0^1 {sds\over r^4(s)}\biggr\rbrace\qquad(5.2.22)} 
\]
\bigskip - Eventually, the third term of (5.2.4) is the more cumbersome one.
It is 
$$
+9\int_0^1\mathrm{d}s\ {\frac{Q_1(R(s))}{r^2(s)}}\int_0^1\mathrm{d}%
s^{\prime}\ [{\widehat r(s)}\!\cdot\!{\widehat r(s^{\prime})}]^2{\frac{%
Q_1(R(s^{\prime}))}{r^2(s^{\prime})}}\eqno(5.2.23) 
$$
An easier way to proceed consists in decomposing the intermediate
integration, on $s^{\prime}$, into 3 pieces : 
\[
\displaylines{\int_0^1{\rm{d}}s'\ [{\widehat r(s)}\!\cdot\!{\widehat r(s')}]^2{Q_1(R(s'))\over
r^2(s')}=(px+qs)^2\int_0^1 {ds'\over r^2(s')}Q_1(s')\cr\hfill+p^2(1-x^2)(p^2-q^2s^2)\int_0^1 {ds'\over r^4(s')}Q_1(s')+2qp^2(1-x^2)(px+qs)\int_0^1 {s'ds'\over r^4(s')}Q_1(s')\qquad(5.2.24)}
\]
\smallskip

- The contribution to the third term of (5.2.4) coming from the 1st term of
(5.2.24) is 
\[
\displaylines{{9\over 2}({p_0}I_3+{q}I'_3-2F_2)\biggl\lbrace -{F_2\over
2}+{1\over 2q}({q+px\over p'^2}-{x\over p})+{p_0\over 2}I_3+{q\over
2}I'_3\cr\hfill -{p_0p^2\over 2}(1-x^2)I_5-{qp^2\over
2}(1-x^2)I'_5\biggr\rbrace\qquad(5.2.25)} 
\]%
\smallskip - The contribution to the third term of (5.2.4) coming from the
second term of (5.2.24) is 
\[
\displaylines{{9\over 2}\left({p^2}(1-x^2)(p_0I_5+qI'_5)+{x\over pq}-{q+px\over qp'^2}-{F_2}\right)\biggl\lbrace -{1\over p'^2}+({px}-{p_0\over 2})I_3-{q\over 2}I'_3\cr\hfill
+{p^2\over 2q}ZI_5+qp\left({p}(1-x^2)+{x\over 2q}Z\right)I'_5\biggr\rbrace\qquad(5.2.26)}
\]%
\smallskip - The contribution to the third term of (5.2.4) coming from the
third term of (5.2.24) is 
\[
\displaylines{{9\over 2}\left(p^2(1-x^2)\left(I_3-p^2I_5+({Z\over 2}-qpx)I'_5\right)-{q+px\over p'^2}+pxF_2\right)\cr\hfill
\times\biggl\lbrace{1\over q}({1\over p'^2}-{1\over p^2})+ I_3+p(xp_0-p)I_5+{Z\over 2}I'_5\biggr\rbrace\qquad(5.2.27)}
\]%
\medskip This shows how incredibly complicated is the exact calculation of
an angular function like $W_{2}^{(2)}(P,P^{\prime })$. 
\subsection{{Collinear singularities} }

\par\bigskip
Let us recall Eq.(3.7) where the counterpart $%
W_{2}^{(2)}(P,P^{\prime })$ of (5.2.2) is now substituted for the whole $%
W^{(2)}(P,P^{\prime })$ of (3.6). The corresponding contribution to $\mathrm{%
Im}\ \Pi _{R}^{(\star ,\star ;2)}(Q)$ one has to examine is 
\[
\displaylines{ \pi e^2m^4\int {{\rm d}^4P\over
(2\pi)^4}(1-2n_F(p_0))\sum_{s,s'=\pm 1}\ \biggl\lbrace -2\pi\beta^{(c)}_s(P)\beta^{(p)}_{s'}(P')W^{(2)}_2(P,P')\cr\hfill 
+\left(\alpha_s(P)\beta^{(p)}_{s'}(P')+\alpha_{s'}(P')\beta^{(p)}_s(P)\right)\left(-i\ {\rm
{disc}}_{p_0}W^{(2)}_2(P,P')\right)\biggr\rbrace\qquad(5.2.28)} 
\]%
Because it is simpler, we begin with analyzing the second term in
the curly bracket of (5.2.28). In the original derivation of the hot QCD collinear singularity problem, it is this term which was responsible for a logarithmic singularity, \cite{[5]}.  

\par\medskip

Out of $W_{2}^{(2)}(P,P^{\prime })$, or (5.2.4), and over the integration range (3.16), all contributions lead to integrals of form
\[
\displaylines{C^{st}\sum_{s,s'}\!\int^q {p^2{\rm d}p\over
(2\pi)^2}\int_{-p}^{p} {{\rm{d}}p_0\over 2\pi}(1-2n_F(p_0))\  \alpha_s(p_0,p)\cr\hfill\times\ \int_{-1}^{p_0/p} {\rm{d}}x\ \delta\left(q+p_0-\omega_{s'}(p'(x))\right)\ (\omega_{s'}^2(p'(x))-{p'^2(x)})\ \frac{1}{(1-x^2)^a}\  {\rm{disc}}_{p_0}\ {H(q,p,p_0;x) }\qquad(5.2.29)}
\]
where (3.11) has been used, and where the power $a$ is in the set $\{0,1,2\}$. Likewise, a function $H(q,p,p_{0};x)$ stands for any of the
functions that can be identified out of Eqs.(5.2.8), (5.2.9),
(5.2.10)-(5.2.15,), (5.2.16), and (5.2.25)-(5.2.27).   Then, inspection shows that the functions generically denoted by $H(q,p,p_{0};x)$ can be decomposed into products of form 
\[
\displaylines{H(q,p,p_0;x)={{{\rm{Pol}}}}\ (p_0;q,p,x)\times {1\over
(2Q\!\cdot\! P )^k}\cr\hfill\times {1\over (p_0^2-2xp_0p+p^2)^b}\times
(\ln{p_0+p\over p_0-p})^c\times (\ln{p'_0+p'\over
p'_0-p'})^{c'}\qquad(5.2.30)} 
\]%
where $\mathrm{Pol}\ (p_{0};q,p,x)$, a polynomial in $p_{0}$, admits a
Taylor series expansion in $x$, and where the integer powers $%
k,b,c,c^{\prime }$ are such that 
$$
0\leq k\leq 2\ ,\ \ \ 0\leq b\leq 4\ ,\ \ \ 0\leq c,c^{\prime }\leq 2\eqno%
(5.2.31) 
$$%

\smallskip In view of the decomposition (5.2.30), the discontinuity in $%
p_{0} $ of any function $H(q,p,p_{0};x)$ splits into a sum of four terms,
any of them proportional to one only of the following list of discontinuities 
$$
\mathrm{disc}_{p_{0}}\ \mathrm{Pol}\ (p_{0};q,p,x)=0\eqno(5.2.32) 
$$%
$$
\mathrm{disc}_{p_{0}}\ {\frac{1}{2Q\!\cdot \!P}}=-2i\pi \delta (2Q\!\cdot
\!P)\eqno(5.2.33) 
$$%
$$
\mathrm{disc}_{p_{0}}\ {\frac{1}{(2Q\!\cdot \!P)^{2}}}=2i\pi \delta ^{\prime
}(2Q\!\cdot \!P)\eqno(5.2.34) 
$$%
$$
\mathrm{disc}_{p_{0}}\ {\frac{1}{p_{0}^{2}-2xp_{0}p+p^{2}}}=-2i\pi \delta
(p_{0}^{2}-2xp_{0}p+p^{2})\eqno(5.2.35) 
$$%
$$
\mathrm{disc}_{p_{0}}\ {\frac{1}{(p_{0}^{2}-2xp_{0}p+p^{2})^{b}}}=-2i\pi {%
\frac{(-1)^{b-1}}{(b-1)!}}\delta ^{(b-1)}(p_{0}^{2}-2xp_{0}p+p^{2})\eqno%
(5.2.36) 
$$%
$$
\mathrm{disc}_{p_{0}}\ (\ln {\frac{p_{0}+p}{p_{0}-p}})^{c}=-ci\pi \Theta
(-P^{2})(\ln {\frac{p_{0}+p}{p_{0}-p}})^{(c-1)}\eqno(5.2.37) 
$$

\par\medskip (i) -The first case, (5.2.32), is trivial.

\par\smallskip (ii) -Terms proportional to the second and third
discontinuities, (5.2.33) and (5.2.34), give zero because of the
incompatibility of $\delta (2Q\!\cdot \!P)$ and $\delta \left( p_{0}^{\prime
}-\omega _{s^{\prime }}(p^{\prime }(x))\right) $.
\par\smallskip (iii) -Terms proportional to the fourth and fifth discontinuities,
(5.2.35) and (5.2.36). At real $p_{0}$-energies, the $\delta
(p_{0}^{2}-2xp_{0}p+p^{2})$ -constraint is satisfied at $x=+1$ where $%
p_{0}=p $, and at $x=-1$ where $p_{0}=-p$, the latter case excluded by
(3.16). Now, a cogent argument, approximation-free and valid at $x=\pm 1$%
, is the following: At $x=\pm 1$, one has $p^{\prime }(x=\pm 1)=q\pm p$,
and so 
$$
\delta (q+p_{0}-\omega _{s^{\prime }}(p^{\prime }))=\delta (q\pm p-\omega
_{s^{\prime }}(q\pm p))\eqno(5.2.38) 
$$%
that has no support in the integration range (and beyond) because, for all $%
s^{\prime }=\pm 1$, the relation $\omega _{s^{\prime }}(q\pm p)-(q\pm p)>0$
holds, in virtue of Fig.3. The two constraints are incompatible, and the corresponding
contributions are zero.
\par\smallskip
 (iv) -Terms proportional to the last discontinuity, (5.2.37) involve both a $%
\Theta (-P^{2})$ and a $\delta (p_{0}^{\prime }-\omega _{s^{\prime
}}(p^{\prime }(x)))$ distribution: They turn out to be identical to the terms related to the first piece of the curly bracket of (5.2.28) that can be analyzed now.
 
\par\bigskip

The first piece in the curly bracket of (5.2.28) requires more care. One can start from an expression similar to (5.2.29),
\[
\displaylines{C^{st}\sum_{s,s'}\!\int^q {p^2{\rm d}p\over
(2\pi)^2}\int_{-p}^{p} {{\rm{d}}p_0\over 2\pi}(1-2n_F(p_0))\  \beta^{(c)}_s(p_0,p)\cr\hfill\times\ \int_{-1}^{p_0/p} {\rm{d}}x\ \delta\left(q+p_0-\omega_{s'}(p'(x))\right)\ (\omega_{s'}^2(p'(x))-{p'^2(x)})\frac{H(q,p,p_0;x)}{(1-x^2)^a}\qquad(5.2.39)}
\]%
with the same set of functions $H(q,p,p_0;x)$ as defined in (5.2.30). In this way, {\it{potential}} collinear singularities are emphasized, as terms proportional to $1/(1-x^2)^a$, with $a\in \{1,2\}$.
For example, such is the case of integrals $I_3$, $I^{\prime}_3$, $I_5$ and $%
I^{\prime}_5$, all of them able to generate collinear singularities at $%
x=\pm 1$, $p_0=\pm p$.  
\par
Clearly, a closer inspection of $H(q,p,p_0;x)$-functions is in order, and more to the point, a regrouping of terms proportional to the potentially dangerous factors of $1/(1-x^2)^a$. 
\par
- Then, one finds that the integrals $I_3$ and $I^{\prime}_3$ are not on the order of $1/(1-x^2)$, but are regular functions of $x$ at $x=\pm 1$, 
$$
\{I_3, I'_3\}_{|_{x=\pm 1}}= {\cal{O}}\left((1\mp x)^0\right)+\dots\eqno(5.2.40)
$$where the dots stand for higher orders in $(1\mp x)$.
\par
- Likewise, and thanks to the very same compensations as those at work in the case of $I_3$ and $I^{\prime}_3$, one finds that $I_5$ and $I^{\prime}_5$ are not on the order of $(1-x^2)^{-2}$, but at worse, on the order of $(1-x^2)^{-1}$,
$$
\{I_5, I^\prime_5\}_{|_{x=\pm 1}}={\cal{O}}\left((1\mp x)^{-1}\right)+ {\cal{O}}\left((1\mp x)^0\right)+\dots\eqno(5.2.41)
$$
 The functions $I_5$ and $I^{\prime}_5$ also depend on the integrals (5.2.20) and (5.2.21) which, like $I_3$ and $I^{\prime}_3$, are regular functions of $x$, at $x=\pm 1$. In the case of (5.2.20) for example, one gets,
 $$\displaylines{\frac{1}{2qp^2(1-x^2)}\left({\frac{px+q}{%
{p^\prime}^2}}-{\frac{x}{p}} +qF_2\right)_{|_{x=\pm 1}}=\frac{1}{4qp^2}\frac{1}{1\mp x}\biggl\lbrace\frac{\pm p+q}{(q\pm p)^2}-\frac{\pm1}{p}\cr\hfill+\frac{q}{p(p\pm q)}+{\cal{O}}(1\mp x)+\dots=0+{\cal{O}}(1\mp x)+\dots\biggr\rbrace\qquad(5.2.42)}
$$
The same applies to (5.2.21), and another similar example of potential collinear singularity compensation will be given below.
\par
- Eventually,  a third useful property is that the four combinations  
$$
pI_n+qxI^{\prime}_n\ ,\ \ \ pxI_n+qI^{\prime}_n\ ,\ \ \ n\in\{3,5\}\eqno%
(5.2.43) 
$$
are able to decrease by one unit the power $a$ of any $1/(1-x^2)^a$%
- contributions appearing in $I_n$ and $I'_n$ (at $n=3$, combinations (5.2.43) are on the order of $(1-x^2)$, in view of (5.2.40)).
\par\medskip
One is now in a position so as to analyze the contributions to (5.2.39) of any of the five terms composing $W^{(2)}_2(P,P')$. For the previous form of (5.2.39), it may be more convenient now, to substitute the expression
\[
\displaylines{C^{st}\sum_{i=1}^5\sum_{s,s'}\!\int^q {p^2{\rm d}p\over
(2\pi)^2}\int_{-p}^{p} {{\rm{d}}p_0\over 2\pi}(1-2n_F(p_0))\  \beta^{(c)}_s(p_0,p)\cr\hfill\times\ \int_{-1}^{p_0/p} {\rm{d}}x\ \delta\left(q+p_0-\omega_{s'}(p'(x))\right)\ (\omega_{s'}^2(p'(x))-{p'^2(x)})\ {T^{(i)}(q,p,p_0;x)}\qquad(5.2.44)}
\]%
where the ${T^{(i)}(q,p,p_0;x)}$ denote the five contributions displayed in (5.2.8), (5.2.9), (5.2.10), (5.2.16) and (5.2.25)-(5.2.27).
\par
- $T^{(5)}$, the fifth term of (5.2.4), given in (5.2.10) and the array of coefficients (5.2.11)-(5.2.15), is a linear combination of regular functions of $x$, $p_0$ and $p$ over the full integration range (3.16), and its contribution to (5.2.44) is singularity free.
\par
- $T^{(2)}$, the second term of (5.2.4), given in (5.2.8), appears singular in the collinear regime of $x=\pm 1$. But it is not so, and at $x=-1$ (as well as at $x=1$) the right hand side of (5.2.8) behaves like
$$
\displaylines{\frac{F_0(x=-1)}{2pq(1+x)}\left(\ln\frac{p'_0+p'}{p'_0-p'}+\ln\frac{p_0+p}{p_0-p}-\ln\frac{P'^2}{P^2}\right)={\cal{O}}(\frac{1}{1+x})(\ln\frac{p_0+p}{p'_0-p'})_{|_{x=-1}}+{\cal{O}}\left(({1+x})^0\right)\cr\hfill
={\cal{O}}(\frac{1}{1+x})\left(\ln\frac{p_0+p}{q+p_0-(q-p)}\right)+{\cal{O}}\left(({1+x})^0\right)={\cal{O}}\left(({1+x})^0\right)\qquad(5.2.45)}
$$
and therefore, its integration over $x$ is collinear singularity free.
\par
- So is also the contribution to (5.2.44) of $T^{(1)}$, the first term of (5.2.4). As displayed by (5.2.9), in effect, this term is the square power of the previous one. In view of (5.2.45), its integration on $x$ is collinear singularity free either.
\par
- $T^{(4)}$, the fourth term of (5.2.4) is given by (5.2.16), and Eqs.(5.2.40), (5.2.41) guarantee that this term lead to collinear singularity free contributions to (5.2.44).
\par
- $T^{(3)}$, the third term of (5.2.4) is given by (5.2.25), (5.2.26) and (5.2.27). The first part, (5.2.25), leads to a regular contribution in virtue of (5.2.40) and (5.2.41). The second part, (5.2.26), leads to a regular contribution in virtue of (5.2.40) and (5.2.41), and also in virtue of the first combination of (5.2.43) taken at $n=5$. The third part, (5.2.27), leads to a regular contribution in virtue of (5.2.40) and (5.2.41), and also in virtue of the two combinations of (5.2.43) taken at $n=5$; to wit, from the second line:
$$
\dots+p(xp_0-p)I_5+{Z\over 2}I'_5=\dots+p_0(pxI_5+qI^\prime_5)-p(pI_5+qxI^\prime_5)=\dots+{\cal{O}}\left(({1-x^2})^0\right)\eqno(5.2.46)
$$where (5.2.41) has been used.
\par
To summarize, relevant regroupings of the initial $H(q,p,p_0;x)$-functions are able to display a full compensation of all {\it{potential}} collinear singularities (both at $x=-1$ and at $x=1$): Certainly, these fine tuning compensations, taking place among so many terms, do not show up by pure chance, and clearly, they support the reliability of the calculations that are presented here.
\par
 As illustrated in Appendix B, collinear singularities would pass from {\it{potential}} to {\it{actual}} upon integration on $x$, $p_0$ and $p$, and not upon integration on $x$ alone.
\par\medskip 
At this point, an important remark is in order. 
\par\noindent
 In the range (3.16), an inspection of the remaining integrations has not revealed any further difficulties: The angular functions $W^{(1)}$ and $W^{(2)}$ display singular behaviours at the light cone $P^2=0$ (of the logarithmic type for example), that do not compromise the regular character of the full integration over (3.16). Now, in this respect, it matters to emphasize that a complete compensation of potential or actual collinear singularities is of utmost importance. As displayed through Appendix B, in effect, terms of $1/(1-x^2)^a$ do not yield any collinear singularity {\it{as such}}. Instead, out of the remaining $p_0$ and $p$-integrations, potential collinear singularities generate further logarithmic and power-law singularities, as well as products of them. 
\par
What is more, these further fake singularities can be proven to receive no screening/removal at all from an improved $HTL$-effective action resumming {\it{asymptotic thermal masses}} along both bosonic and fermionic lines, \cite{[8]}. This very unusual circumstance, fully understandable though, is demonstrated in Appendix B.
\par
 These examples therefore, are highly suggestive of the crucial importance of potential/actual collinear singularity compensations, if any. Missing the completeness of collinear singularity compensations, results into severe further troubles:  As suggested in Appendix B, the collinear-enhancement mechanism and the related loop-expansion breaking, are very likely nothing else than some of these troubles.

\section{Conclusion}

\bigskip The hot QCD collinear singularity problem had to be revisited
entirely, and were it not for the tedious calculations this revisitation
requires, the task could have been achieved sooner.

\smallskip As we have seen in effect, the thorough evaluation of entwined
angular averages is very complicated and a lot of patient checkings
is needed. This is the more so as, to our knowledge (and ability) at least,
no mathematical program is really able to yield the full results of Section
5.2. Getting them however, is the price to be paid in order to fix
definitely that 16 years old issue (experience shows in effect, that
attempts at guessing the essential features of such complicated objects as
those entwined angular averages, are doomed to failure).

\medskip Our results can be summarized as follows.

\smallskip

\textbf{-} In the first place, having proceeded, within the correct
sequence, to a most careful analysis of the $1$- and $2$- effective vertex
diagrams relevant to the soft photon emission rate out of a Quark-Gluon
Plasma at thermal equilibrium, we claim that the corresponding emission rate
is singularity-free. The hot QCD collinear singularity problem simply
doesn't exist, and in textbooks, should no longer be presented as a serious
obstruction to the Resummation Program.

\smallskip As pinned up in \cite{[11]}, the 1994's- famous divergent result, 
\cite{[5]}, is due to erroneous manipulations due to the fuzzy distinction made
in our formalisms, between the prescriptions of \textit{discontinuity} and 
\textit{imaginary part}. Whereas the latter commutes with an integration
process, by integration's linearity, the former does not, in
general, because it is defined by a limiting procedure, \cite{[11]}. Now, the prevalence of the discontinuity prescription over the imaginary part one has been advocated in Ref.\cite{[19]}.
\par
\textbf{-} Appendix B has revealed instructive aspects. To summarize, if for some reason, a complete compensation of (potential or actual) collinear singularities is missed, then, the same drawbacks occur, as encountered  by the improved effective perturbation theory: 
\par
- Resummed asymptotic thermal masses, bosonic and fermionic, do not provide enough screening, and, due to power-law collinear-induced singularities, a full leading order emission rate calculation requires higher order diagrams. 
\par
- Moreover, the required extra diagrams may clearly depend on the regulators that are choosen in order to quantify the collinear-induced singularities of the original diagrams.  An unavoidable arbitrariness is thus introduced in the emission rate leading order completion, supposing under control that extra diagrams are determined at the exclusion of any others. 
\par
- And last but not least, extra diagrams are also expected to compensate for original singularities, \cite{[21]}. But we have seen here, how very peculiar to the diagram under consideration, are the collinearly generated singularities. Now, infrared/collinear cancellations between diagrams of different topologies, \cite{[21]}, in a non-abelian context, what is more, \cite{[22]}, is a highly non-trivial conjecture, if not an exceptional one: If that possibility can be thought of as reliable, at least so long as the stronger infrared singularities are concerned, \cite{[23]}, there is no guarantee whatsoever that it could be so in the case of sub-leading ones; quite on the contrary, \cite{[24]}.
\par
- It is therefore suggested that all of the above long known difficulties, express an incomplete compensation of initial collinear singularities, and nothing else. This is the more likely so, as the calculations presented here are able, among so many terms, to exhibit a fine tuning compensation of all of the possible collinear singularities: A cogent enough result, which can not happen just by chance.
\par
Accordingly, right from hot QCD first principles, the finite contributions that remain provide us with a sound, reliable basis for a complete leading order estimate of the soft photon emission rate out of a QGP at thermal equilibrium, \cite{[25]}. This perspective should be of interest in view of RHIC and LHC experimental runs.

\appendix
\section{} 

The second line of Eq.(4.5) defines $F(p_{0})$, the function

$$
F(p_{0})=\sum_{s^{\prime }=\pm 1}\ \int_{-1}^{\frac{p_{0}}{p}}\mathrm{d}x\ {%
\frac{\delta \left( \omega _{s^{\prime }}(p^{\prime }(x)-q-p_{0}\right) }{%
2q(\omega _{s^{\prime }}(p^{\prime }(x))-(q+px))}}\ (\omega _{s^{\prime
}}^{2}(p^{\prime }(x))-{p^{\prime }{}^{2}(x)})(1-s^{\prime }{\frac{\omega
_{s^{\prime }}(p^{\prime }(x))}{p^{\prime }(x)}})\eqno(A.1) 
$$%
where the denominator has been expressed as $2Q\!\cdot \!P^{\prime }$. We
recall that $p^{\prime }{}^{2}(x)=p^{2}+2pqx+q^{2}$ and ${\vec{q}}\!\cdot \!{%
\vec{p^{\prime }}}=q(q+px)$. In order to see the regular character of $%
F(p_{0})$ in a neighborhood of $p_{0}=p$, one may expand the constraint of $%
\delta \left( \omega _{s^{\prime }}(p^{\prime }(x)-q-p_{0}\right) $, 
$$
\delta \left( \omega _{s^{\prime }}(p^{\prime }(x)-q-p_{0}\right) =\delta
\left( \omega _{s^{\prime }}(p^{\prime }(x)-(q+p)\right) -(p-p_{0}){\frac{%
\mathrm{d}}{\mathrm{d}(q+p)}}\delta (q+p-\omega _{s^{\prime }}(p^{\prime
}(x)))+..\eqno(A.2) 
$$%
where the dots stand for higher order corrections in $(p-p_{0})$, and obtain 
\[
\displaylines{F(p_0)=\sum_{s'=\pm 1}\ \biggl\lbrace\int_{-1}^{+1} {\rm{d}}x\ {\delta\left(q+p-\omega_{s'}
(p'(x))\right)}\ (1-s'{q+p
\over p'(x)})\cr\hfill -(p-p_0){{\rm{d}}\over {\rm{d}}(q+p)}\int_{-1}^{+1} {\rm{d}}x\ { \delta(q+p-\omega_{s'}(p'(x))) }\ (1-s'{q+p
\over p'(x)})+ ..\biggr\rbrace\qquad(A.3)} 
\]%
Note that instead of $P^{2}<0$, which in addition to $p_{0}>-p$, (3.16),
would also preclude any risk of potentially singular behavior at $p_{0}=p$,
one allows for $P^{2}\leq 0$ in view of the step-function $\Theta (-P^{2})$
appearing in (3.10). This is equivalent to $P^{\prime }{}^{2}-2Q\!\cdot
\!P^{\prime }\leq 0$, which at $p_{0}^{\prime }=q+p_{0}=\omega _{s^{\prime
}}(p^{\prime }(x))$ is guaranteed, provided the inequality $q-p\leq \omega
_{s^{\prime }}(p^{\prime }(x))\leq q+p$ be satisfied. By picking up a value
of $x$ smaller than 1 (at $x=1$ in effect, any term in the expansion (A.3)
would be zero because of the argument of Eq.(5.2.38), the latter
inequality allows for the constraint of $\delta (q+p-\omega _{s^{\prime
}}(p^{\prime }(x)))$ to have a non-empty support in the integration domain
(3.16).

\section{}

In this appendix, it is assumed that a full compensation of collinear singularities is not obtained, so that in (5.2.39), some function $H(q,p,p_{0};x)$ remains, hereafter denoted by ${\cal{H}}(q,p,p_{0};x)$, whose behaviour at $x=-1$ does not compensate for the potentially dangerous factors of $1/(1-x^2)^a$. What happens then?

\par \smallskip In
order to examine the behavior of (5.2.39) in the collinear regime of $%
x\simeq -1$, one can rely on the expansion \cite{[6]} 
$$
\omega _{s^{\prime }}(p^{\prime }(x))\simeq m\left( 1+s^{\prime }{\frac{%
p^{\prime }(x)}{3m}}+{\cal{O}}({\frac{p^{\prime }}{m}})^2\right) \ ,\ \ \ x\simeq -1%
\eqno(B.1) 
$$%
This expansion makes sense provided that $p^{\prime }(x)/m<<1$, a condition
that is met at $x\simeq -1$, in view of (3.16) and in view also, of the relatively
narrow phase-space extension of the Resummation Program, \cite{[15], [16]}.
Without prejudice to our concern, $(B.1)$ allows us to replace in
(5.2.39), the factor $\omega _{s^{\prime }}^{2}(p^{\prime }(x))-{p^{\prime
}{}^{2}(x)}$ by the constant $m^{2}$, because one has, \cite{[6]}, 
$$
{\frac{\omega _{s^{\prime }}^{2}(p^{\prime })-p^{\prime }{}^{2}}{2m^{2}}}%
\simeq {\frac{1}{2}}+s^{\prime }{\frac{p^{\prime }}{m}}\eqno(B.2) 
$$%
\smallskip \noindent The contribution to (5.2.39) of the collinear regime $%
x\simeq -1$ therefore reads as 
\[
\displaylines{\sim C^{st}{9m^2\over q}\sum_{s,s'}\!\int^q {p{\rm d}p\over
(2\pi)^2} \left({3qp\over 2(q-p)}\right)^a\cr\hfill\times\int_{p_{01}}^{+p} {{\rm{d}}p_0\over 2\pi}(1-2n_F(p_0))\  \beta^{(c)}_s(p_0,p)\ {\cal{H}}(q,p,p_0;x_0)(q+p_0-m)\ \left({1\over p_0-p_{01}}-{1\over p_0-p_{02}}\right)^a\qquad(B.3)}
\]%
where the two zeros $p_{0i}$ are 
$$
p_{01}\simeq m-p-{\frac{2}{3}}(q-p)\ ,\ \ \ p_{02}\simeq m-p-{\frac{4}{3}}(q-p)\eqno%
(B.4) 
$$%
and where, bearing on the angle selected by the constraint $\delta \left( \omega _{s^{\prime }}(p^{\prime }(x)-q-p_{0}\right)$, the condition
$$
-1\leq x_0=-1+\frac{9(p_0-p_{01})(p_0-p_{02})}{2qp}
\eqno(B.5)
$$
restricts the original $p_0$-range to the interval $[p_{01},+p]$.
\par\medskip
 The ensuing integrations do not exist in the rigorous
mathematical sense because of a pole at $p_0=p_{01}$, and another one at $p=q$, both induced by a potential collinear singularity at $x=-1$. As is often the case at non-zero temperature, \cite{[16], [17]}, extra regularizations must be supplied. Let it be done by shifting the pole at $p_{01}$ a small amount of $\lambda$, and the pole at $p=q$, a small amount of $\delta q$. Then, in $(B.3)$, two values of $a$ come into play:
\par (i) At $a=1$, and in the limit of vanishing regulators, $\lambda=0$ and $\delta q=0$, two logarithmically divergent contributions come out, on the order of
$$
{\cal{O}}\left(\ln\frac{\delta q}{q}\right)+ {\cal{O}}\left(\ln\frac{ q}{\lambda}\right)
\eqno(B.6)
$$

\par (ii) At $a=2$, to the two previous singular behaviours, $(B6)$, one must add singular contributions on the strength of $$
{\cal{O}}\left(\ln\frac{\delta q}{q}\times\ln\frac{ q}{\lambda}\right)+ {\cal{O}}\left(\frac{ 1}{\lambda}\right)+{\cal{O}}\left(\frac{ 1}{\delta q}\right)+{\cal{O}}\left(\frac{1}{\delta q}\times\ln\frac{ q}{\lambda}\right)+{\cal{O}}\left(\frac{1}{\lambda}\times\ln\frac{q}{\delta q}\right)
\eqno(B.7)
$$
It is worth remarking that these singular behaviours are generated by the $x=-1$-collinear regime: Whereas a genuine collinear singularity, as such, does not appear, a {\it{potentially}} singular collinear behaviour is at the origin of the {\it{actual}} singular terms developed by the remaining integrations on $p_0$ and $p$.
\par
\par (iii) An amazing feature revealed by this calculation is worth emphasizing. If one proceeds to improve the bosonic and fermionic $HTL$-effective actions in the sense of Ref.\cite{[8]}, providing gluon and quark fields with asymptotic thermal masses $m_\infty$ and $M_\infty$ respectively, then, no change is brought to the above situation. Contrarily to usual expectations, the singularities of $(B.6)$ and $(B.7)$ receive no screening from a resummation of asymptotic thermal masses.  
\par\medskip
This can be seen as follows. Asymptotic thermal masses will affect the effective quark propagators, the ${}^\star S_R(P)$ of (2.2), by substituing to (2.3) an improved version of the thermal self energy,
 $$
 (2.3)\longrightarrow m^2\ \frac{2}{\pi^2}\int_0^\infty {\rm{d}}\alpha\ \frac{\alpha\ e^\alpha}{e^{2\alpha}-1}\int {\frac{\mathrm{d}{\widehat K}}{4\pi}}\left( {\frac{{%
\rlap
/ \!\widehat K} }{{\ \widehat K}\!\cdot\! P+\frac{{\rm{d}}m}{\alpha}}}+{\frac{{%
\rlap
/ \!\widehat K} }{{\ \widehat K}\!\cdot\! P-\frac{{\rm{d}}m}{\alpha}}}\right)\eqno(B.8)
$$
where, 
$$
{\rm{d}}m=\frac{m_\infty^2-M_\infty^2}{2T},\ \ m_\infty^2=\frac{g^2T^2N}{6},\ \ \ M_\infty^2=\frac{g^2T^2C_F}{8}\eqno(B.9)
$$
And likewise, the effective photon-quark-quark vertex, the $\Gamma _{\mu }^{HTL}(P_{\alpha },Q_{\beta },P_{\delta }^{\prime })$ of (2.5) is improved in a similar way, \cite{[8]},
$$
C^{st}\int_0^\infty {\rm{d}}\alpha\ \frac{\alpha\ e^\alpha}{e^{2\alpha}-1}\int {\frac{\mathrm{d}{\widehat K}}{4\pi}}\left( {\frac{{%
\rlap
/ \!\widehat K}\ {\widehat {K}^\mu} }{({\ \widehat K}\!\cdot\! P+\frac{{\rm{d}}m}{\alpha})({\ \widehat K}\!\cdot\! P'+\frac{{\rm{d}}m}{\alpha})}}+ {\frac{{%
\rlap
/ \!\widehat K}\ {\widehat {K}^\mu} }{({\ \widehat K}\!\cdot\! P-\frac{{\rm{d}}m}{\alpha})({\ \widehat K}\!\cdot\! P'-\frac{{\rm{d}}m}{\alpha})}} \right)\eqno(B.10)
$$

 \par\medskip
	 Surprisingly enough, in the latter case, factors of $(1-x^2)^{-a}$, at $a=1$ and $a=2$, are left the same as at ${\rm{d}}m=0$, and collinear singularities at $x=-1$ receive no screening. As inspection shows in effect (Section V.B), this is so because those singular factors come exclusively from the $r^{-3}(s)$, $r^{-4}(s)$ and $r^{-5}(s)$ pieces of the various functions to be integrated over $s\in [0,1]$: One has $r^{2}(s)=p^{2}+2pqxs+q^{2}s^{2}$, (5.1.12), and Tables show that the ensuing integrations come out proportional to inverse powers of $\Delta=4p^2q^2-(2pqx)^2$, \cite{[13]}. Now, the function $r^2(s)$ itself comes from the scalar product ${\vec{r}}(s)\!\cdot\! {\vec{r}}(s)$, where the vector ${\vec{r}}(s)={\vec{p}}+s{\vec{q}}$ is clearly unaffected by a shift of $p_0$ to $p_0\pm{\rm{d}}m/\alpha$, followed by an average over $\alpha$.  
\par 
Note that a full calculation only, was able to reveal such a fate, so as the related peculiar nature of the ensuing collinear singularities.
 
\par\medskip
In the former case, quasi-particle poles, $\omega_{ s}(p)$ of Eq.(3.8), are solutions of
$$
p_0-sp-{\frac{m^{2}}{2p}}\left(
(1-s{\frac{p_{0}}{p}})\ln {\frac{p_{0}+p}{p_{0}-p}}+2s\right)=0
\eqno(B.11)
$$
and with $(B.8)$, will now become solutions of
$$
p_0-sp-s\frac{m^2}{p}-\frac{1}{2}\biggl\lbrace\biggl\langle\sum_{\eta=\pm 1}\frac{m^{2}}{2p}
(1-s{\frac{p_{0}+\eta\frac{dm}{\alpha}}{p}})\ln {\frac{p_{0}+p+\eta\frac{dm}{\alpha}}{p_{0}-p+\eta\frac{dm}{\alpha}}}\biggr\rangle\biggr\rbrace=0
\eqno(B.12)
$$
where the notation $\langle\dots\rangle$ has been introduced as a shorthand to mean
$$\biggl\langle F(\alpha)\biggr\rangle=\frac{2}{\pi^2}\int_0^\infty {\rm{d}}\alpha\ \frac{\alpha\ e^\alpha}{e^{2\alpha}-1}F(\alpha)\eqno(B.13)
$$
For this new equation to admit a new solution in the integration range of (3.16), ${\widehat{\omega}}_{ s}(p)$, the new terms composing it cannot be an order of magnitude bigger than the remaining ones, and in view of $(B.9)$, this observation imposes $g/\alpha<1$, that is, $\alpha$ cannot be smaller than $g$. 
 Then, considering the $P'$-fermionic line, relevant to the crossed possibilities of (3.10), after some algebra, it is possible to re-write $(B.12)$ as,
$$
p'_0-s'p'(x)-s'\frac{m^2}{2p'}(1-s'\frac{p'_0}{p'})\ln\frac{p'_0+p'}{p'_0-p'}-s'\frac{m^2}{p'}\left(1+\frac{1}{2}s'(1-s'\frac{p'_0}{p'})\langle\frac{1}{\alpha}\rangle\frac{p'_0}{m}\!\cdot\!\frac{dm}{m}\right)+\dots=0\eqno(B.14)
$$
where the dots stand for higher orders in a small parameter development, the parameter $p{\rm{d}}m/\alpha P^2$. At $p_0\pm p={\cal{O}}(gT)$, this small parameter is on the order of ${\rm{d}}m/\alpha (p_0\pm p)\simeq {\cal{O}}(g/6\alpha)$, with $g/\alpha<1$. Note that in $(B.14)$, because of $\alpha>g$, we have now a slight modification of the average introduced in $(B.13)$,
$$\biggl\langle \frac{1}{\alpha}\biggr\rangle={\rm{C^{st}}}(g)\int_g^\infty {\rm{d}}\alpha\ \frac{\alpha\ e^\alpha}{e^{2\alpha}-1}\ \frac{1}{\alpha}=\frac{1}{2}{\rm{C^{st}}}(g)\ln \frac{1}{g}=(\frac{2}{\pi^2}\ln \frac{1}{g})(1+{\cal{O}}(g))\eqno(B.15)
$$
A comparison of $(B.14)$ to $(B.11)$ shows that the former, whith respect to the same equation taken at ${\rm{d}}m=0$, is modified an amount of relative magnitude
$$\frac{s'}{2}(1-s'\frac{p'_0}{p'})\langle\frac{1}{\alpha}\rangle\frac{p'_0}{m}\!\cdot\!\frac{dm}{m}=-\frac{1}{\pi^2}{\sqrt{\frac{2}{3}}}\frac{p'_0}{m}\ g\ln g\equiv \varepsilon({\rm{d}}m)={\cal{O}}(g\ln g)\eqno(B.16)
$$
which can still be taken as a small enough quantity.
\par
Denoting by $\omega'_{ s'}(p')$ the solutions to $(B.14)$ taken at ${\rm{d}}m=0$, this suggests to look for solutions to $(B.14)$ under the form of 
$${\widehat{\omega'}}_{ s'}(p')=\omega'_{ s'}(p')+\delta\omega'_{ s'}(p')\eqno(B.17)
$$and to analyze the consequence on the pole-location, coming from the new inherited constraint of $\delta (q+p_0-{\widehat{\omega'}}_{ s'}(p'(x)) )$. Then, provided that the condition
$$\frac{\delta \omega'_{s'}}{\omega'_{s'}-p'}<<1\eqno(B.18)
$$is satisfied, one finds
$$
\delta\omega'_{s'}(p')=-2\varepsilon({\rm{d}}m)\frac{s'm^2(\omega'-s'p')}{-p'(\omega'-s'p')-s'm^2+2(\omega'-s'p')^2\frac{m^2}{\omega'^2-p'^2}}
\eqno(B.19)
$$where $\omega'$ is a shorthand for $\omega'_{ s'}(p'(x))$. Since $\omega'_{ s'}(p'(x))$ complies with the expansion $(B.1)$ for $x$ in a neighbourhood of $-1$, this expansion can be used in $(B.19)$ so as to get
$$\delta\omega'_{s'}(p')=-\varepsilon({\rm{d}}m)\left(\frac{2s'm}{2-s'}+{\cal{O}}(\frac{p'}{m})\right)\eqno(B.20)
$$
and we note that $(B.20)$ complies with $(B.18)$,
$$
\frac{\delta \omega'_{s'}}{\omega'_{s'}-p'}\simeq {\cal{O}}(\varepsilon({\rm{d}}m))\eqno(B.21)$$
Since $\delta\omega'_{s'}(p')$ is proportional to $m$, the new constraint, $\delta (q+p_0-{\widehat{\omega'}}_{ s'}(p'(x)) )$, will amount to re-define the pole at $p_{01}$,
$$p_{01}\longrightarrow {\widehat{p_{01}}}\simeq m\left(1-\frac{2s'}{2-s'}\varepsilon({\rm{d}}m)\right)-p-{\frac{2}{3}}(q-p)\eqno(B.22)
$$
with, obviously, the same conclusions $(B.6)$ and $(B.7)$ as at ${\rm{d}}m=0$. 
\par\bigskip
What can be learned out of this example? Apparently, that if a complete compensation of potential/actual collinear singularities is missed, then further singularities develop, and that a resummation of thermal asymptotic masses does not bring enough screening, to say the less. Also, since some of the generated singularities are power-law, what shouldn't come as a surprise (see Ref.\cite{[8]}), it becomes obvious that, depending on the scale of the adopted regulators (the $\lambda$ and $\delta q$), higher number of loop diagrams will be found to be on the same orders of magnitude as elementary ones.
\par
Needless to emphasize that these difficulties are of course very similar to those which are known to plague the $HTL$- improved effective perturbation theory, \cite{[9]}. Note also that, depending on the scale of the adopted regulators, the extra diagrams that will become necessary to complete the full soft photon emission rate leading order, will differ .. and in any case, will be hoped to cancel out the original singularities, \cite{[21]}, ..
\par

\end{document}